\documentclass[twocolumn,tighten]{aastex62}

\graphicspath{{./}{figures/}}

\shorttitle{The Angular Momentum of the CGM in TNG100}
\shortauthors{DeFelippis et al.}

\begin{document}

\title{The Angular Momentum of the Circumgalactic Medium in the TNG100 Simulation}

\correspondingauthor{Daniel DeFelippis}
\email{d.defelippis@columbia.edu}

\author{Daniel DeFelippis}
\affiliation{Department of Astronomy, Columbia University, 550 West 120th Street, New York, NY 10027, USA}

\author{Shy Genel}
\affiliation{Center for Computational Astrophysics, Flatiron Institute, 162 Fifth Avenue, New York, NY 10010, USA}

\author{Greg L. Bryan}
\affiliation{Department of Astronomy, Columbia University, 550 West 120th Street, New York, NY 10027, USA}
\affiliation{Center for Computational Astrophysics, Flatiron Institute, 162 Fifth Avenue, New York, NY 10010, USA}

\author{Dylan Nelson}
\affiliation{Max-Planck-Institut f\"ur Astrophysik, Karl-Schwarzschild-Str. 1, D-85741 Garching, Germany}

\author{Annalisa Pillepich}
\affiliation{Max-Planck-Institut f\"ur Astronomie, K\"onigstuhl 17, D-69117 Heidelberg, Germany}

\author{Lars Hernquist}
\affiliation{Harvard-Smithsonian Center for Astrophysics, 60 Garden Street, Cambridge, MA 02138, USA}

\begin{abstract}

We present an analysis of the angular momentum content of the circumgalactic medium (CGM) using TNG100, one of the flagship runs of the IllustrisTNG project. We focus on Milky Way$-$mass halos ($\sim 10^{12} \; M_{\odot}$) at $z=0$ but also analyze other masses and redshifts up to $z=5$. We find that the CGM angular momentum properties are strongly correlated with the stellar angular momentum of the corresponding galaxy: the CGM surrounding high-angular momentum galaxies has a systematically higher angular momentum and is better aligned to the rotational axis of the galaxy itself than the CGM surrounding low-angular momentum galaxies. Both the hot and cold phases of the CGM show this dichotomy, though it is stronger for colder gas. The CGM of high-angular momentum galaxies is characterized by a large wedge of cold gas with rotational velocities at least $\sim1/2$ of the halo's virial velocity, extending out to $\sim 1/2$ of the virial radius, and by biconical polar regions dominated by radial velocities suggestive of galactic fountains; both of these features are absent from the CGM of low-angular momentum galaxies. These conclusions are general to halo masses $\lesssim 10^{12} \; M_{\odot}$ and for $z \lesssim 2$, but they do not apply for more massive halos or at the highest redshift studied. By comparing simulations run with alterations to the fiducial feedback model, we identify the better alignment of the CGM to high-angular momentum galaxies as a feedback-independent effect and the galactic winds as a dominant influence on the CGM's angular momentum. 

\end{abstract}


\section{Introduction} \label{sec:intro}

The circumgalactic medium (CGM), loosely defined as the nonstellar baryonic material filling the region outside of galaxies but within their dark matter halos, is thought to contain significant amounts of baryonic mass, and many recent observational and computational studies focus on it as a key to better understanding galaxy formation and evolution. The CGM can be both a vessel of cosmological gas accretion and a reservoir of gas ejected out of galaxies by feedback (see \citealp{Tumlinson17} and references therein). The combination of accreting, ejected, and quasi-hydrostatic components, which continuously interact, mix, and get replenished, potentially shows signatures in the kinematics of the CGM, with various degrees of rotation and complex velocity structure. Recent studies using both cosmological simulations \citep[e.g.][]{Ford14} and zoom-in simulations \citep[e.g.][]{Angles-Alcazar17,Hafen19} generally find many different origins and evolutionary histories of gas in the CGM, supporting this picture. Furthermore, the fact that galaxies and their CGM are physically connected to each other and exchange mass suggests that certain properties of the two may be, in general, correlated \citep[see, e.g.,][]{Bland-Hawthorn17}. Similarly, the diversity of observed galaxies suggests that the CGM may be diverse as well, whether because accreting gas in the CGM affects the evolution of the galaxy, or gas ejected from the galaxy affects the evolution of the CGM.

Observations of the CGM have generally supported these conclusions. Through studies of both small samples \cite[e.g.][]{Bouche16,Rahmani18,Lochhaas19,MartinD19} and large samples \cite[e.g.][]{Bordoloi14,Liang14,Turner14,Werk14,Kacprzak15,Schroetter16,Turner17,McQuinn18,Burchett19,Pointon19} of absorption lines, as well as studies using emission lines \cite[e.g.][]{Martin15,Martin16}, the CGM has been found to contain a mass of baryons comparable to that of its associated galaxy, composed of gas in many different ionization and dynamical states, indicative of many channels of formation.

Close to the plane of the galactic disk, the velocity of this material is often consistent with being corotating with the galaxy (first seen in \citealp{Barcons95}; more recent works include \citealp{Ho17,Zabl19}). These observations are limited by the nature of the observational technique, which integrates gas absorption along line-of-sight ``pencil beams'' through the CGM and can combine gas at different radii and in different dynamical states \citep[e.g.][]{Kacprzak19,Ng19}. Gravitational lensing can allow a quasar to probe multiple discrete locations \citep[e.g.][]{Chen14} or a continuous region \citep[e.g.][]{Lopez18,Lopez20} in the same CGM: such special cases are also consistent with corotation. However, in all of these cases, the real 3D motions of gas in the CGM are reduced to 1D line-of-sight velocities, meaning direct measurements of more fundamental vector quantities like angular momentum are very difficult to achieve.  Therefore, even though there is observational evidence for high-angular momentum gas in the Milky Way's (MW's) CGM (probed by quasar sightlines out to $\sim 80 \; \rm{kpc}$ into the halo in \citealp{Hodges-Kluck16}) and in specific higher-redshift MW analogs (probed by Ly$\alpha$ up to $\sim40 \; \rm{kpc}$ from the galaxy in \citealp{Prescott15}), the angular momentum of gas in the CGM is rarely directly studied.  

Angular momentum has long been considered an important quantity in galaxy structure and evolution. It is predicted theoretically to originate from tidal torquing by the cosmic web at high redshifts \citep{Peebles69,Fall80,Mo98} and has been shown to strongly correlate with galaxy morphology, both in observations \citep{Fall13,Cortese16,Swinbank17} and in large cosmological simulations \citep{Genel15,Teklu15,Zavala16}. Angular momentum contained in galaxies and halos has been measured in cosmological simulations including Illustris \citep{Rodriguez-Gomez17,Zjupa17} and EAGLE \citep[e.g.][]{Lagos17,Stevens17,Oppenheimer18}, in zoom-in simulations of individual halos \citep{El-Badry18,Garrison-Kimmel18}, and using analytic models \citep{Pezzulli17,Sormani18}. However, the nature of the relationship between the angular momentum of the galaxy and the angular momentum of the CGM is not yet clear. 

A well-established result from both zoom-in and large-scale cosmological simulations is that galaxies can eject low-angular momentum gas into the CGM while also accreting higher-angular momentum gas from the CGM that eventually can form stars \citep{Brook11,Brook12,Ubler14,Christensen16,DeFelippis17,Grand19}. In particular, \cite{Brook12} and \cite{DeFelippis17} found that much of the gas that forms stars by $z=0$ has been ejected into and reaccreted from the CGM successively, each time with incrementally more angular momentum. Other studies have found similar links that relate the misalignment of the galaxy and halo to properties of accreting gas \citep{Roskar10}, outflowing gas \citep{Tenneti17}, and satellite galaxies \citep{Shao16} in the halo. Furthermore, \cite{Stewart13} and \cite{Stewart17} have found, using zoom-in simulations, that the CGM can develop a cold extended disk of high-angular momentum gas from cosmological accretion. These results all suggest that the CGM could generally be a source and reservoir not just of gas, in general, but of angular momentum for galaxies. To determine whether this is the case would require measuring the angular momentum in the CGM of a large sample of realistic galaxies, which has not yet been done for a large cosmological simulation.

It is with this motivation in mind that we seek, as a first step, to characterize the angular momentum of the CGM for a large population of galaxies from the TNG100 simulation to determine what, if any, systematic properties appear. In Section \ref{sec:methods}, we describe the IllustrisTNG simulation suite and our angular momentum calculations in detail, as well as define key properties of our CGM sample. In Section \ref{sec:results} we describe our main results. In Section \ref{sec:discussion}, we discuss the implications and possible physical origins of our results, and we summarize in Section \ref{sec:summary}. We plan, in future papers, to follow up this theoretically based work in two ways: by determining to what extent our conclusions are supported by observations, and by tracing the overall and detailed evolution of gas throughout the CGM using the IllustrisTNG simulations.

\section{Methods} \label{sec:methods}

\subsection{Simulations}

This work makes use of the TNG100 box of the IllustrisTNG simulation suite \citep{Marinacci18,Naiman18,Nelson18,Pillepich18,Springel18}, which utilizes the moving-mesh code \textsc{Arepo} \citep{Springel10,Weinberger19} to evolve a periodic $\approx (111 \; \rm{Mpc})^{3}$ box from cosmological initial conditions down to $z=0$. It has a baryonic mass resolution of $1.4\times10^6 \; M_{\odot}$ per cell. Two forms of feedback are included: (1) galactic winds launched using energy released from evolving stellar populations \citep{Pillepich18a}, and (2) energy ejections from active galactic nuclei (AGNs) that occur in two modes corresponding to high and low accretion rates onto the black hole \citep{Weinberger17}. IllustrisTNG is based on the original Illustris simulation suite \citep{Vogelsberger13,Vogelsberger14a,Vogelsberger14b,Genel14} and notably improves upon its feedback prescriptions to be more in line with the observational constraints for both high-mass \citep{Nelson18,Pillepich18} and low-mass \citep{Pillepich18a} galaxies.

\subsection{Analysis}

Our halo selection is made based on the virial mass $M_{\rm vir}$, calculated with a spherically averaged overdensity criterion \citep{Bryan98} for objects identified with the friends-of-friends algorithm \citep{Davis85}. However, we wish to distinguish between the contribution of satellite galaxies and the contribution of the main central galaxy. Thus, for each halo, we select (i) gas cells from the central subhalo as calculated by the \textsc{subfind} algorithm \citep{Springel01}, (ii) gas contained in satellite subhalos identified by the same algorithm, and (iii) gas contained in the halo ``fuzz'' that is geometrically part of the halo but not bound to any subhalo. We then define ``smooth'' gas as all gas bound to the central subhalo or part of the halo fuzz ((i)and (iii)) and satellite gas as all gas bound to satellite subhalos (ii). Finally, we define the smooth (satellite) component of the CGM as all smooth (satellite) gas that is outside of a sphere centered on the most bound particle contained in the galaxy with radius equal to twice the stellar half-mass radius. The gas inside this sphere, though we do not consider it in this paper, we call the interstellar medium (ISM). \cite{Pillepich19} found in TNG50 that the extent of the star-forming gas (i.e. gas with a number density $n > 0.13 \; \rm{cm}^{-3}$) relative to the stellar half-mass radius depends on both halo mass and redshift. We find this as well in TNG100, but the impact on the CGM mass is minimal: at $z=0$, star-forming gas is on average $<2\%$ of the CGM by mass for all halo masses considered in this paper, and it is $<10\%$ for all $z\leq2$. We have also tried defining the CGM with other geometric and/or mass cuts but settled on twice the stellar half-mass radius due to its relative simplicity and so as to mimic how galaxy sizes are measured observationally with stellar light. Regardless, we generally find that our results are not sensitive to the precise definition we choose here. We also further divide the smooth CGM into ``hot'' and ``cold'' phases based on a threshold at half the virial temperature, $T_{\rm vir}$, defined as
\begin{equation} 
T_{\rm vir} = \frac{\gamma-1}{k_B} \frac{GM_{\rm vir}}{R_{\rm vir}} \mu    
\label{eq:Tvir}
\end{equation}
where $\gamma = 5/3$, and $\mu$ is the mean molecular weight of the gas in the halo. For halos of masses $\sim10^{11} \; M_{\odot}$, $\sim10^{12} \; M_{\odot}$, and $\sim10^{13} \; M_{\odot}$, $T_{\rm vir} $ has average values of $\approx2\times10^5 \; \rm{K}$, $\approx8\times10^5 \; \rm{K}$, and $\approx4\times10^6 \; \rm{K}$ respectively, and the ``cold'' phase comprises on average $\approx80\%$, $\approx66\%$, and $\approx50\%$ of the total CGM gas mass, respectively. Note that by our definition, all gas is either hot or cold.

We calculate the specific angular momentum of the CGM of a galaxy as follows:
\begin{equation} 
\mathbf{j}_{\rm{CGM}} = \frac{1}{M_{\rm{CGM}}}\sum_{i=1}^{N} m_{i}(\mathbf{r}_{i}-\mathbf{r}_{\rm{center}})\times(\mathbf{v}_{i}-\mathbf{v}_{\rm{com}})
\label{eq:jcgm}
\end{equation}
where the summations go over all gas cells included as part of the CGM or of one of its components, as appropriate. $M_{\rm{CGM}} = \sum_{i=1}^{N} m_{i}$ is the total mass of that component, $\mathbf{r}_{\rm{center}}$ is the position of the most bound particle in the halo, and $\mathbf{v}_{\rm{com}}$ is the center-of-mass velocity of the central galaxy, defined as the collection of all of the stars in the central subhalo. We also define the misalignment angle of the CGM with respect to the galaxy as the angle between $\mathbf{j}_{\rm{CGM}}$ and the stellar specific angular momentum vector ($\mathbf{j}_{*}$), where $\mathbf{j}_{*}$ is calculated as in Equation \ref{eq:jcgm} but using the central galaxy's star particles.

\begin{figure}
\fig{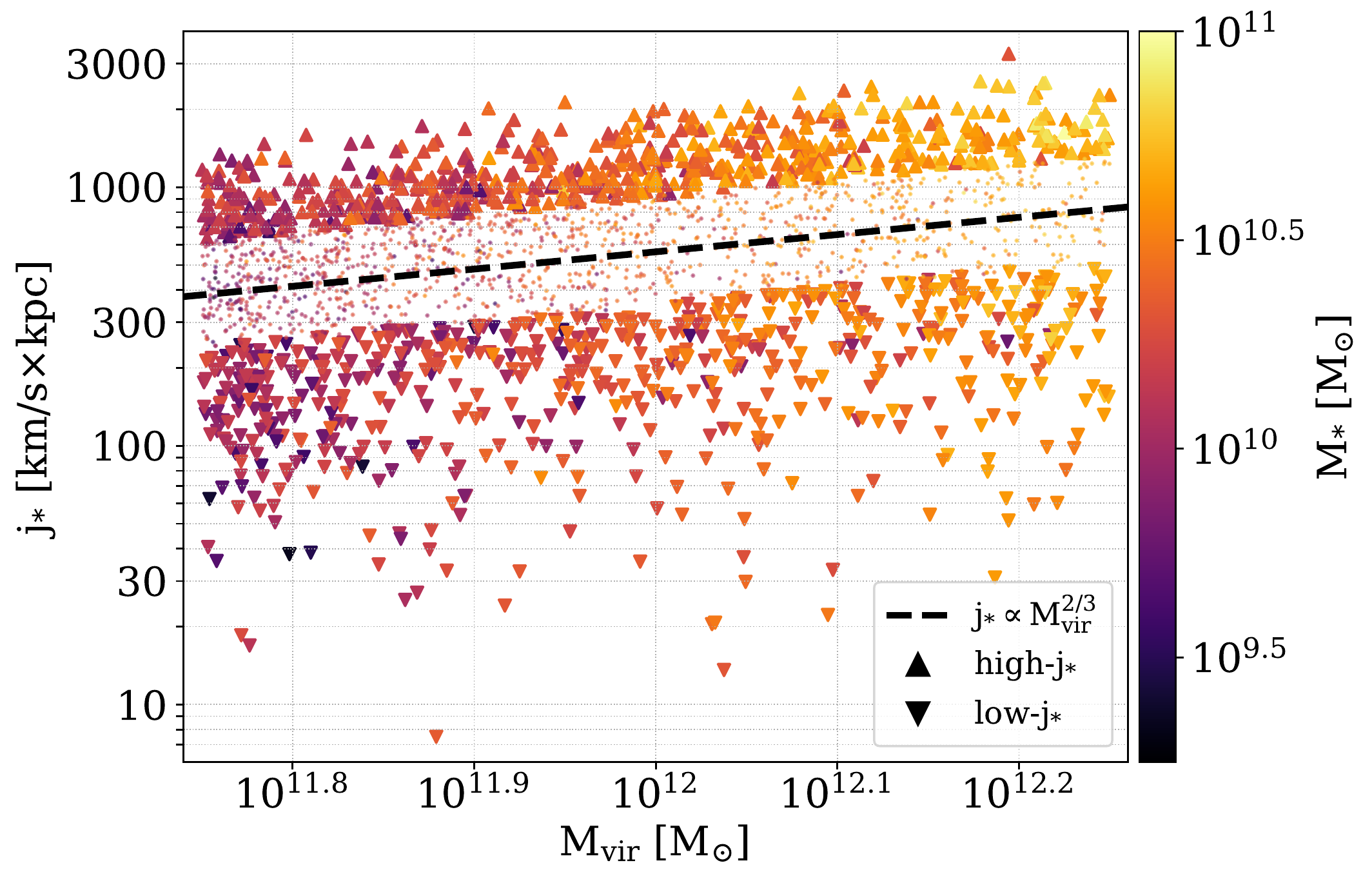}{0.48\textwidth}{}
\caption{Stellar specific angular momentum magnitude vs. the virial mass of TNG100 MW-mass halos at $z=0$. The high-$j_{*}$ and low-$j_{*}$ galaxies as defined in Section \ref{sec:methods} are shown as upward and downward facing triangles, and the two middle quartiles of the $j_{*}$ distribution are shown as smaller circles. Each point is colored by the galaxy's stellar mass. The black dashed line indicates a power law in this plane with a slope of $2/3$.
}
\vspace{0.3cm}
\label{f:jstars}
\end{figure}

Following our previous analyses of baryonic angular momentum in the Illustris simulation \citep{Genel15,DeFelippis17}, we divide halos into two populations based on the stellar specific angular momentum magnitude ($j_{*}$) of the central galaxy (i.e. the central \textsc{subfind} subhalo). For the MW-mass scale, which is our focus, we select all halos with virial masses in the range $10^{11.75} < M_{\rm vir} < 10^{12.25} \; M_{\odot}$ to have as large a sample of as possible. Rather than using the distribution of $j_{*}$ alone, we first normalize $j_{*}$ by $M_{\rm vir}^{2/3}$ and choose the upper and lower quartiles of the distribution of $j_{*}M_{\rm vir}^{-2/3}$, a quantity that is functionally similar to the traditional halo spin parameter $\lambda \propto j_{\rm{halo}}M_{\rm vir}^{-2/3}$. This is necessary for removing any lingering halo mass dependence and ensuring that the two populations have essentially the same mass distributions. We refer to the upper and lower quartiles defined above, which each contain 630 halos, as high-$j_{*}$ and low-$j_{*}$ respectively, and we show the full $j_{*}$ distribution of our halo mass range and highlight the high-$j_{*}$ and low-$j_{*}$ samples in Figure \ref{f:jstars}. We note that the high-$j_{*}$ sample has a lower median black hole mass and higher specific star-formation rate than the low-$j_{*}$ sample by factors of $\approx 0.6$ and $\approx 1.4$, respectively.

\section{Results} \label{sec:results}

In this section, we present the results of our analysis technique as described in Section \ref{sec:methods}. We primarily focus on MW-mass halos at $z=0$ (Section \ref{sec:results1}), and then expand our analysis to higher redshifts (Section \ref{sec:results2}) and other halo masses (Section \ref{sec:results3}). 

\subsection{MW-mass Halos at $z=0$} \label{sec:results1}

\begin{figure}
\fig{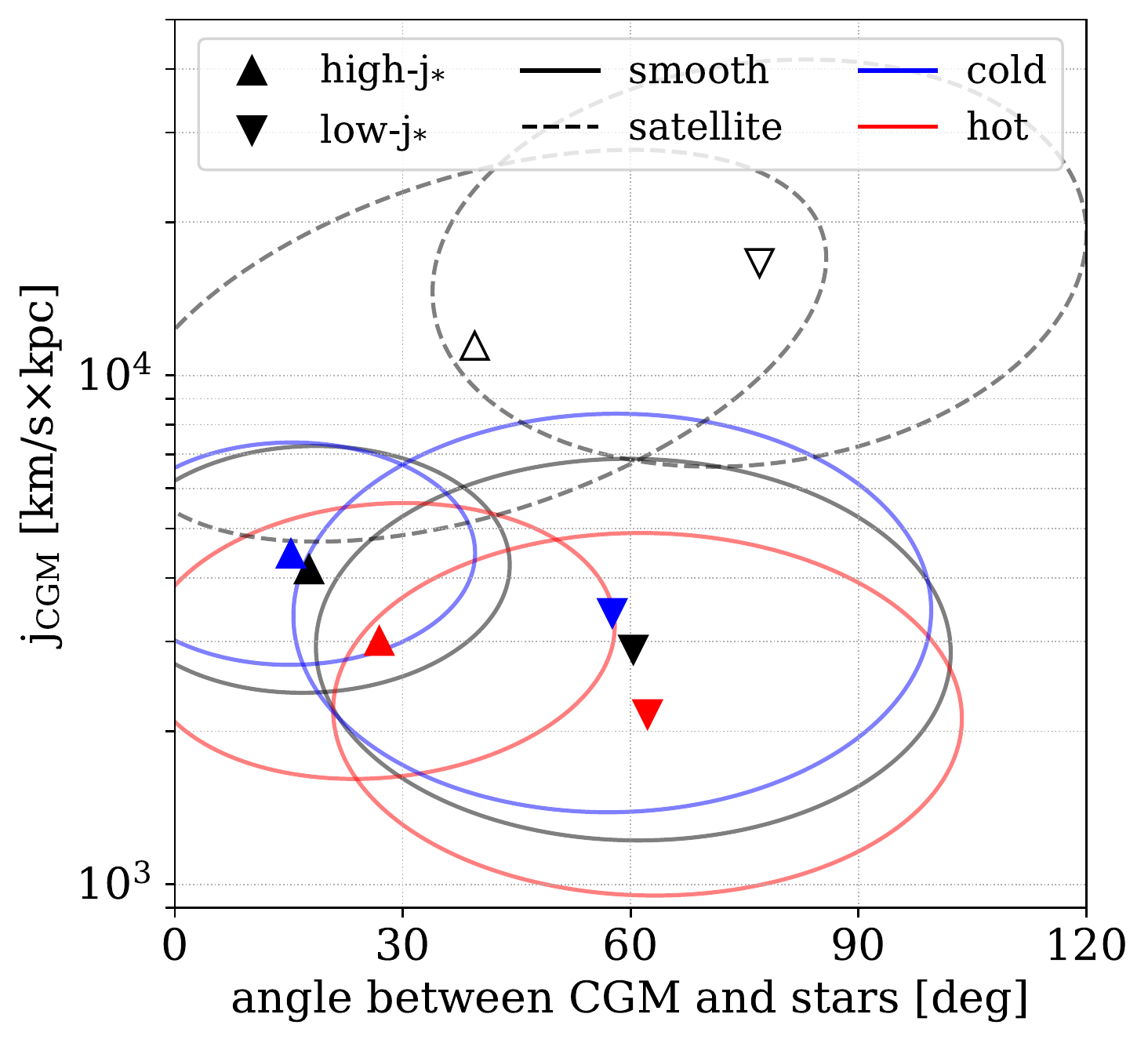}{0.48\textwidth}{}
\caption{
Specific angular momentum magnitude vs. misalignment angle of CGM gas with respect to the stellar angular momentum axis for TNG100 MW-mass halos at $z=0$. The median values are shown for all gas (black) around high-$j_{*}$ galaxies (upward pointing triangles) and low-$j_{*}$ galaxies (downward pointing triangles).  This is shown both for the smooth (i.e. non-satellite) component (solid, filled triangles) and for satellites (dashed, empty triangles). The smooth component is further divided into cold (blue) and hot (red) components based on a temperature threshold of $T_{\rm{vir}}/2$ for each halo. The ellipses surrounding each median point show the corresponding $1\sigma$ scatter of the covariance between the magnitude and misalignment.
}
\vspace{0.3cm}
\label{f:totalj}
\end{figure}

We begin in Figure \ref{f:totalj} by plotting the total specific angular momentum magnitude of the CGM against its misalignment angle relative to the stars for high-$j_{*}$ and low-$j_{*}$ MW-mass halos at $z=0$ as defined in Section \ref{sec:methods}. We also distinguish between the smooth and satellite components as defined above. First, we consider the smooth component shown as the solid black triangles: we see that the entire smooth CGM of high-$j_{*}$ galaxies (upward triangles) is very well aligned to the stars in the galaxy, with a median misalignment angle of only about $15^{\circ}$. Conversely, the smooth CGM of low-$j_{*}$ galaxies (downward triangles) is much more poorly aligned, with a higher misalignment angle by $\sim40^{\circ}$, and a lower specific angular momentum magnitude, by about a factor of $1.5$. By comparing Figures \ref{f:jstars} and \ref{f:totalj}, we can see that the CGM of high-$j_{*}$ galaxies has a $\sim3-4$ times higher specific angular momentum than that of the high-$j_{*}$ galaxies themselves (namely their stellar component), while for low-$j_{*}$ galaxies, this ratio is typically larger, $\sim10$.

Further, Figure \ref{f:totalj} also shows the same quantities but now split into the hot and cold components of the smooth CGM. We see that the $\sim40^{\circ}$ misalignment angle difference and factor of $\sim1.5$ magnitude difference between high-$j_{*}$ and low-$j_{*}$ galaxies is present in each of the cold and hot gas components separately. In detail, the cold and hot components differ (in a similar way for high-$j_{*}$ and low-$j_{*}$ galaxies) such that the cold component around both galaxy types has a slightly higher magnitude and is somewhat better aligned to the stars than the hot component. We found these results to be insensitive to temperature cutoffs anywhere between 0.1 and 1 $T_{\rm vir}$.

Next, we look at the satellite component. We note that, unlike for the smooth component, we do not define a cutoff around the stellar disk of any satellite galaxy, as the entire satellite subhalo is part of the CGM of the central subhalo. Gas in satellites (empty triangles in Figure \ref{f:totalj}) has a much higher specific angular momentum than the smooth CGM$-$about $0.5$ dex for high-$j_{*}$ galaxies and 1 dex for low-$j_{*}$ galaxies$-$and both satellite components are less aligned to the stars than their corresponding smooth components are. However, on average, the smooth component contains an order-of-magnitude more mass, which means the total angular momentum contents of the smooth and satellite components are roughly equal.

\begin{figure*}
\gridline{
\fig{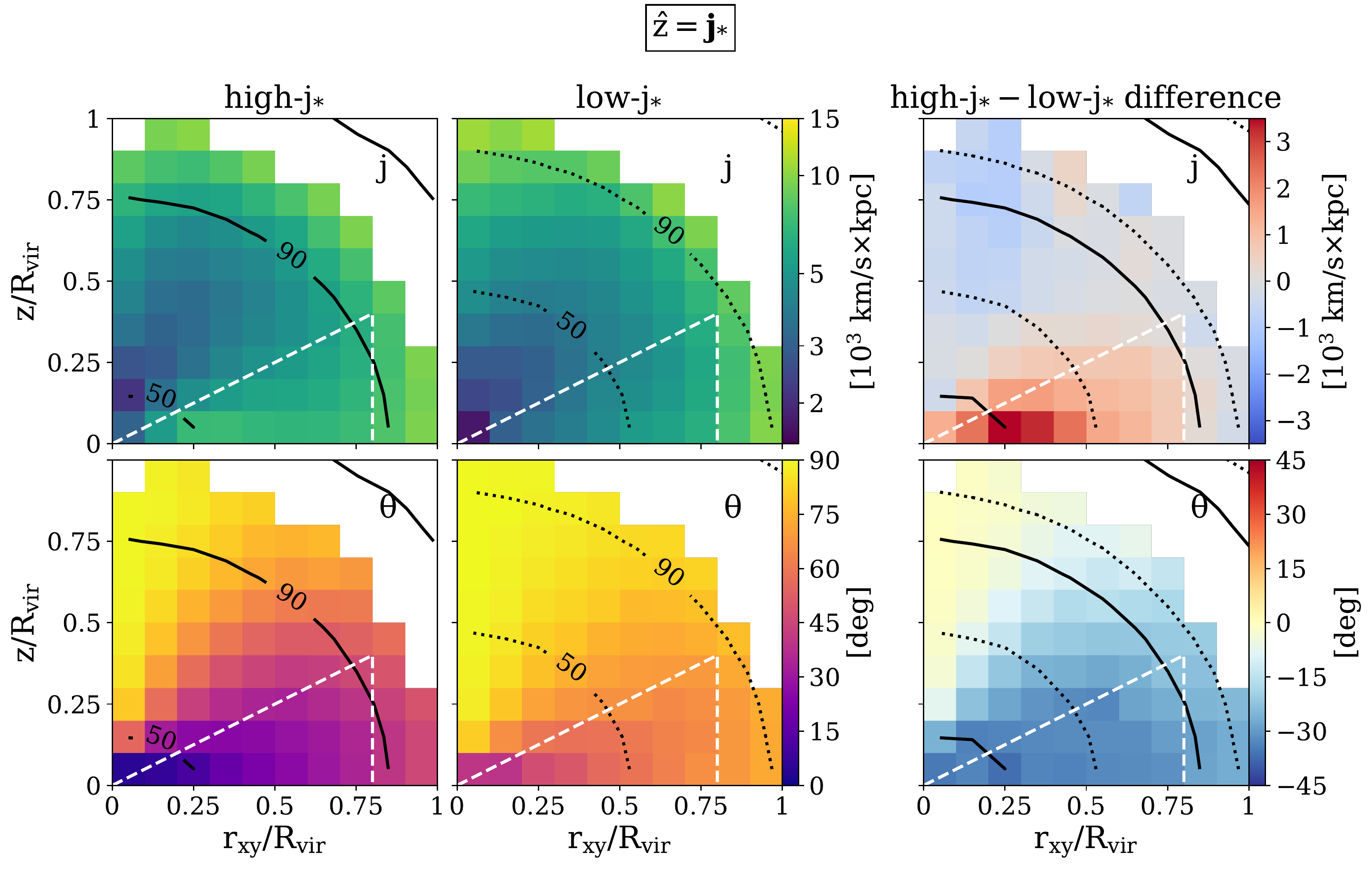}{0.715\textwidth}{}
\hspace{0.5cm}
\fig{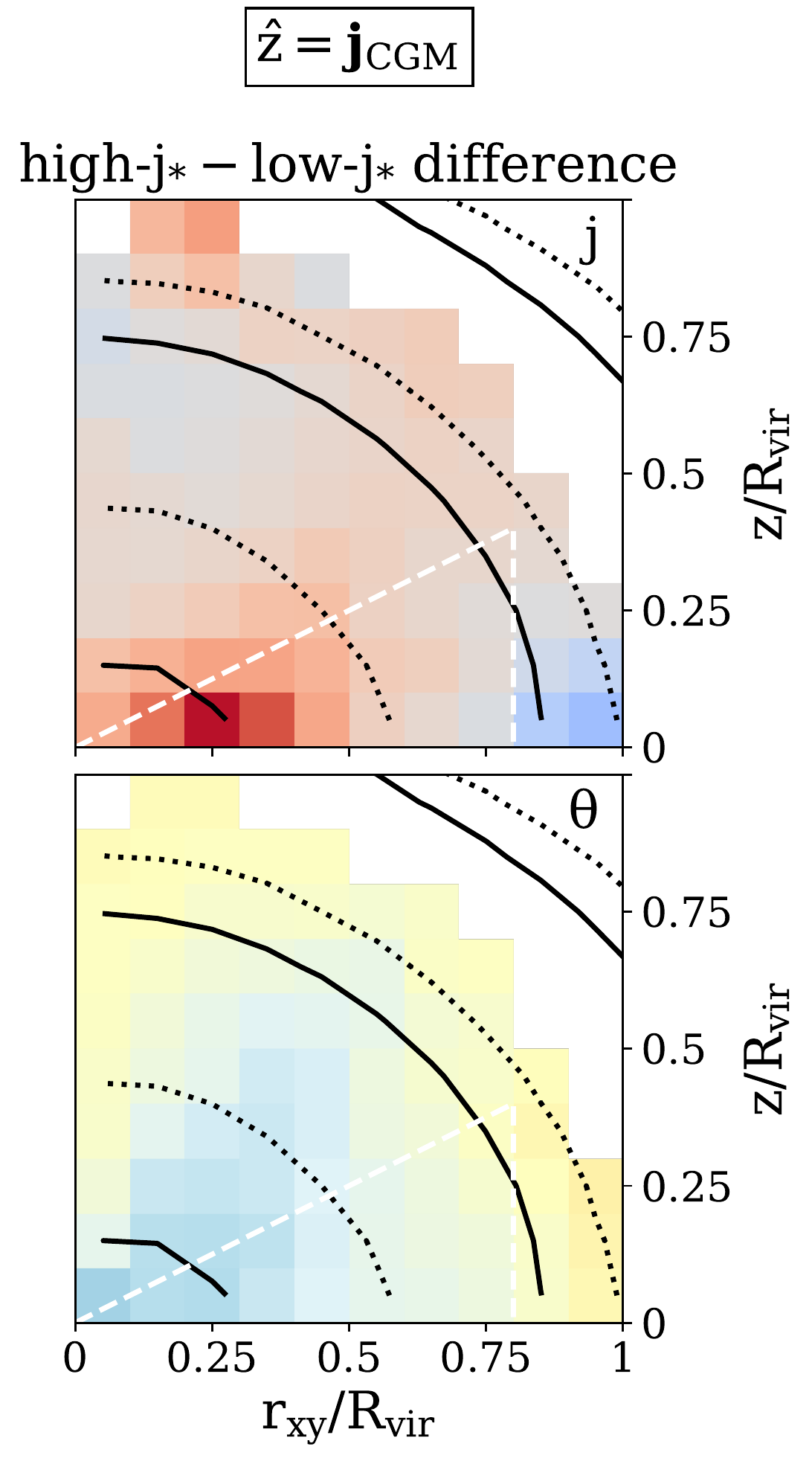}{0.25152\textwidth}{}
}
\caption{
Each panel shows mass-weighted spatial distributions of the cold CGM binned in height ($y$-axis) and cylindrical radius ($x$-axis) for TNG100 MW-mass halos at $z=0$. The first three columns are computed with the $z$-axis pointing in the direction of the stellar angular momentum vector. The top row displays specific angular momentum magnitudes in units of $1000 \; \rm{km} \; \rm{s}^{-1} \; \rm{kpc}$ and the bottom row displays misalignment angles in degrees. The first column shows the actual angular momentum magnitude and misalignment angle of the CGM around high-$j_{*}$ galaxies, and the second column shows the same for the CGM around low-$j_{*}$ galaxies. The third column is simply the difference between the first column and the second column. The final column is calculated in the same way as the third column but with the $z$-axis pointing in the direction of the total angular momentum vector of the CGM, rather than the stars. The black contours in each panel are isodensity contours of cold gas, labeled by the percentage of cold gas mass (50\%, 90\%, and 99\%) they enclose. Rounder (flatter) contours therefore highlight more (less) spherically symmetric density profiles. The white dashed triangle is meant to guide the eye by emphasizing the properties of the gas within a $\sim30^{\circ}$ wedge centered on the plane perpendicular to the axis of rotation. 
}
\label{f:coldjprofile}
\end{figure*}

\begin{figure*}
\gridline{
\fig{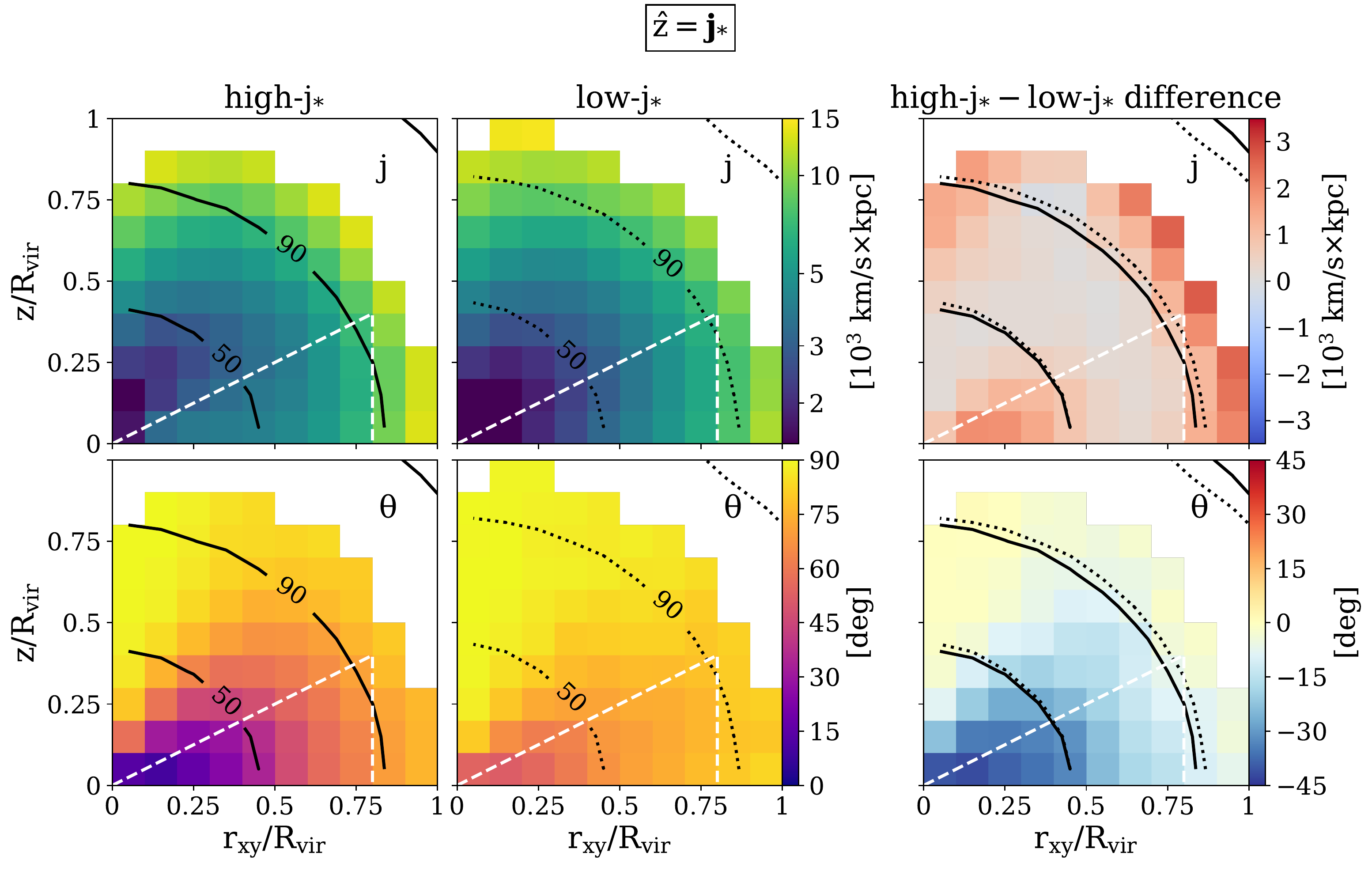}{0.715\textwidth}{}
\hspace{0.5cm}
\fig{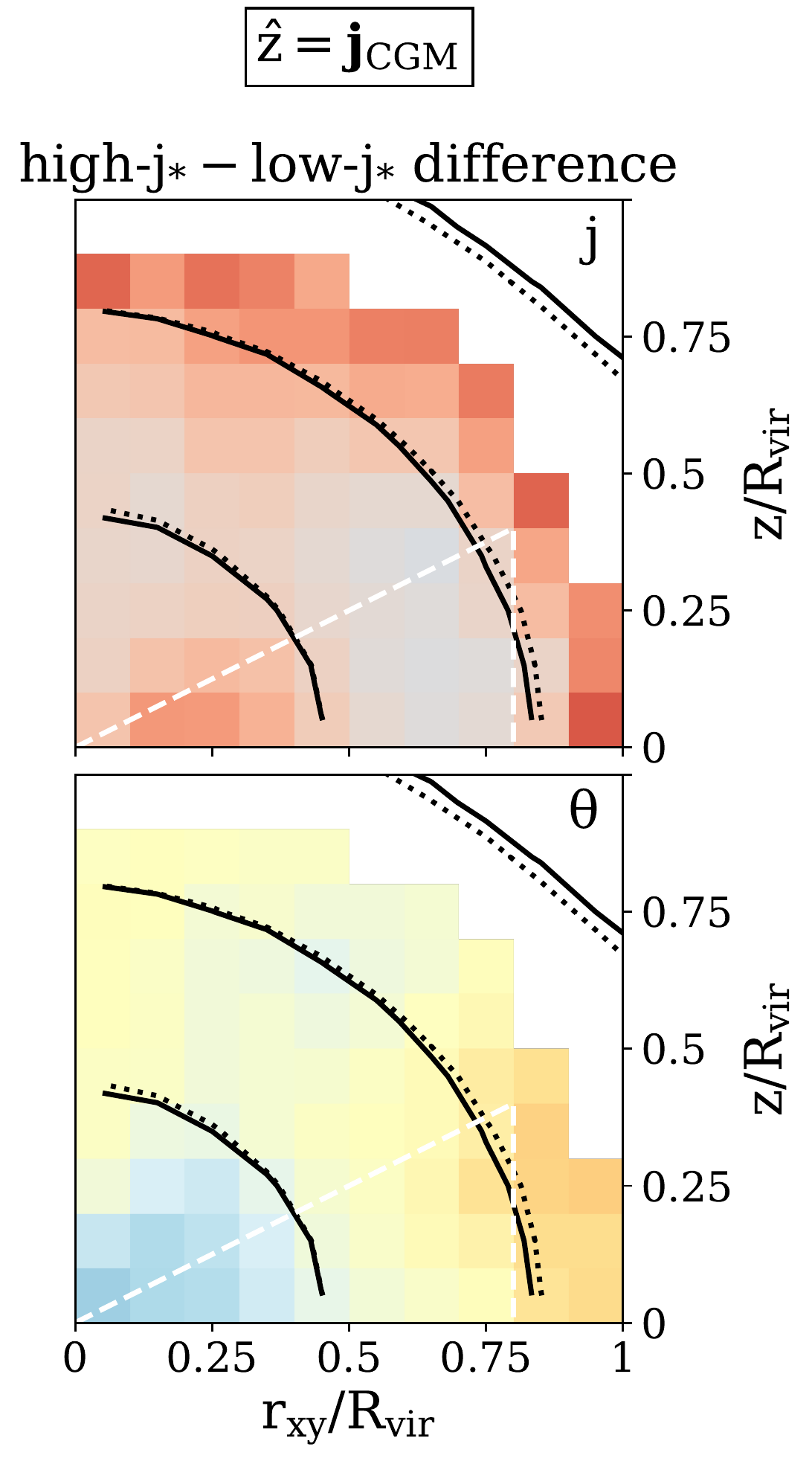}{0.25152\textwidth}{}
}
\caption{
Each panel shows the mass-weighted spatial distributions of the hot CGM for TNG100 MW-mass halos at $z=0$ and plots the same quantity as the corresponding panel in Figure \ref{f:coldjprofile}.
}
\label{f:hotjprofile}
\end{figure*}

In comparing high-$j_{*}$ and low-$j_{*}$ galaxies, we see a smaller misalignment angle with respect to the stars for satellites around high-$j_{*}$ galaxies, just like for the smooth component. But, unlike the smooth component, there is a larger magnitude around the low-$j_{*}$ galaxies. However, there are reasons to believe the same high-$j_{*}$/low-$j_{*}$ split is not very meaningful for the satellite component. First, the mass fraction of the CGM in the satellite component is extremely variable: $\sim20\%$ of MW-mass halos contain no satellite component to their CGM at all, and $\sim2/3$ of them have less than $5\%$ of their CGM mass contained in satellites; still, $\sim1\%$ of halos actually have a majority of their CGM mass in satellites. Second, the scatter of the satellite component as shown in Figure \ref{f:totalj} is quite large compared to that of the smooth component, especially in the misalignment angle. Third, the median misalignment between the satellite component and the smooth component (not shown) is nearly the same for high-$j_{*}$ and low-$j_{*}$ galaxies ($30^{\circ}$ and $40^{\circ}$, respectively) and is notably in between the medians of the high-$j_{*}$ and low-$j_{*}$ smooth component. These results indicate that the angular momentum found in satellites at $z=0$ is generally associated with a small amount of mass at misalignment angles more related to the smooth component than to the galaxy. This can be understood as follows: at $z=0$, the smooth component of the CGM is made up of accreted and ejected gas averaged over all cosmic time, while the satellite component is much more transient and subject to strong variations. In other words, the satellite component is tracing a much shorter timescale of accretion and is thus more weakly related to the angular momentum of the galaxy. Therefore, for the rest of this paper, we will focus on the smooth component of the CGM where there is more gas and where the split in high-$j_{*}$ and low-$j_{*}$ populations is strongest and more physically meaningful, presumably highlighting an important connection between the CGM and the galaxy. In particular, we want to identify the gas that could be driving the difference in the overall properties of the two populations.

In Figure \ref{f:coldjprofile}, we show the distribution of the average angular momentum magnitude (top row) and misalignment angle (bottom row) in different spatial bins of the cold, smooth CGM component. Figure~\ref{f:hotjprofile} shows the same quantities but for the hot, smooth component. Our goal in displaying the CGM this way is to understand which parts of the CGM are significantly different between high-$j_{*}$ and low-$j_{*}$ galaxies and, thus, where the overall angular momentum trends in Figure \ref{f:totalj} come from. Before discussing the results, we provide two minor caveats. First, we only display properties of the gas out to 
a spherical radius of $R_{\rm vir}$ because past this boundary, we find all measured properties to be very noisy. Second, within $R_{\rm vir}$, we only show averages in spatial bins where at least 50\% of the halos have gas in that bin.

Immediately evident in the first two columns of Figures \ref{f:coldjprofile} and \ref{f:hotjprofile}, especially in Figure \ref{f:coldjprofile}, is a high-angular momentum and well-aligned gaseous ``wedge'' centered on the galactic plane and extending to large radii in high-$j_{*}$ galaxies that barely exists in low-$j_{*}$ galaxies. To help guide the eye, such a wedge is outlined with dashed white lines in each panel of Figures \ref{f:coldjprofile} and \ref{f:hotjprofile}. Such a feature is perhaps expected for cold gas, which is generally the more centrally concentrated (as shown by the black isodensity contours, which are labeled with the percentage of mass enclosed) and rotationally supported, but interestingly, the hot gas also shows this feature, albeit to a lesser extent in angular momentum magnitude. This can also be seen quantitatively in the third column, which is simply the difference between the first two columns: the CGM of high-$j_{*}$ galaxies with respect to low-$j_{*}$ galaxies has an angular momentum excess of as much as 3000 $\rm{km \; s^{-1} \; kpc}$ in cold gas and 1000 $\rm{km \; s^{-1} \; kpc}$ in hot gas within this wedge, and it is also $\approx 30^{\circ}$ better aligned to the stars in this wedge, independent of temperature. Outside the wedge region, the differences between the CGM of high-$j_{*}$ and low-$j_{*}$ galaxies are much smaller. We comment here that the existence of this wedge demonstrates that the angular momentum distribution of the CGM is more cylindrically symmetric (i.e. symmetric with respect to the $z$-axis) than spherically symmetric in the halo, which \cite{Bullock01} also found to be true for dark-matter-only simulations of comparable halos.

Finally, we calculate the same spatial distributions for the CGM of high-$j_{*}$ and low-$j_{*}$ galaxies but with the $z$-axis set to the direction of $\mathbf{j}_{\rm{CGM}}$ rather than $\mathbf{j}_{*}$. We note that the misalignment angle in this case is no longer between the CGM and the galaxy but between the local CGM and global CGM. We find that the CGM properties of high-$j_{*}$ galaxies hardly change (as expected, since the galaxy and the CGM are rather well aligned, see Figure \ref{f:totalj}), but the CGM properties of low-$j_{*}$ galaxies do, resulting in the difference maps shown in the fourth columns of both Figures \ref{f:coldjprofile} and \ref{f:hotjprofile}. Compared to those in the third columns, the misalignment angle difference maps show much smaller values, meaning the local-to-global CGM misalignment angles of the CGM of high-$j_{*}$ and low-$j_{*}$ galaxies are similar. Therefore, large misalignment differences in the third columns of Figures \ref{f:coldjprofile} and \ref{f:hotjprofile} are due to the overall misalignment of the CGM of low-$j_{*}$ galaxies to the galaxy, rather than the internal properties of the CGM around those galaxies. However, while the misalignment angle difference is significantly lessened, the angular momentum magnitude difference hardly changes, and there is still a nonzero misalignment angle difference at $r_{xy} < 0.5 R_{\rm vir}$ and $z < 0.5 R_{\rm vir}$ in both the hot and cold phases. 

\begin{figure}
\fig{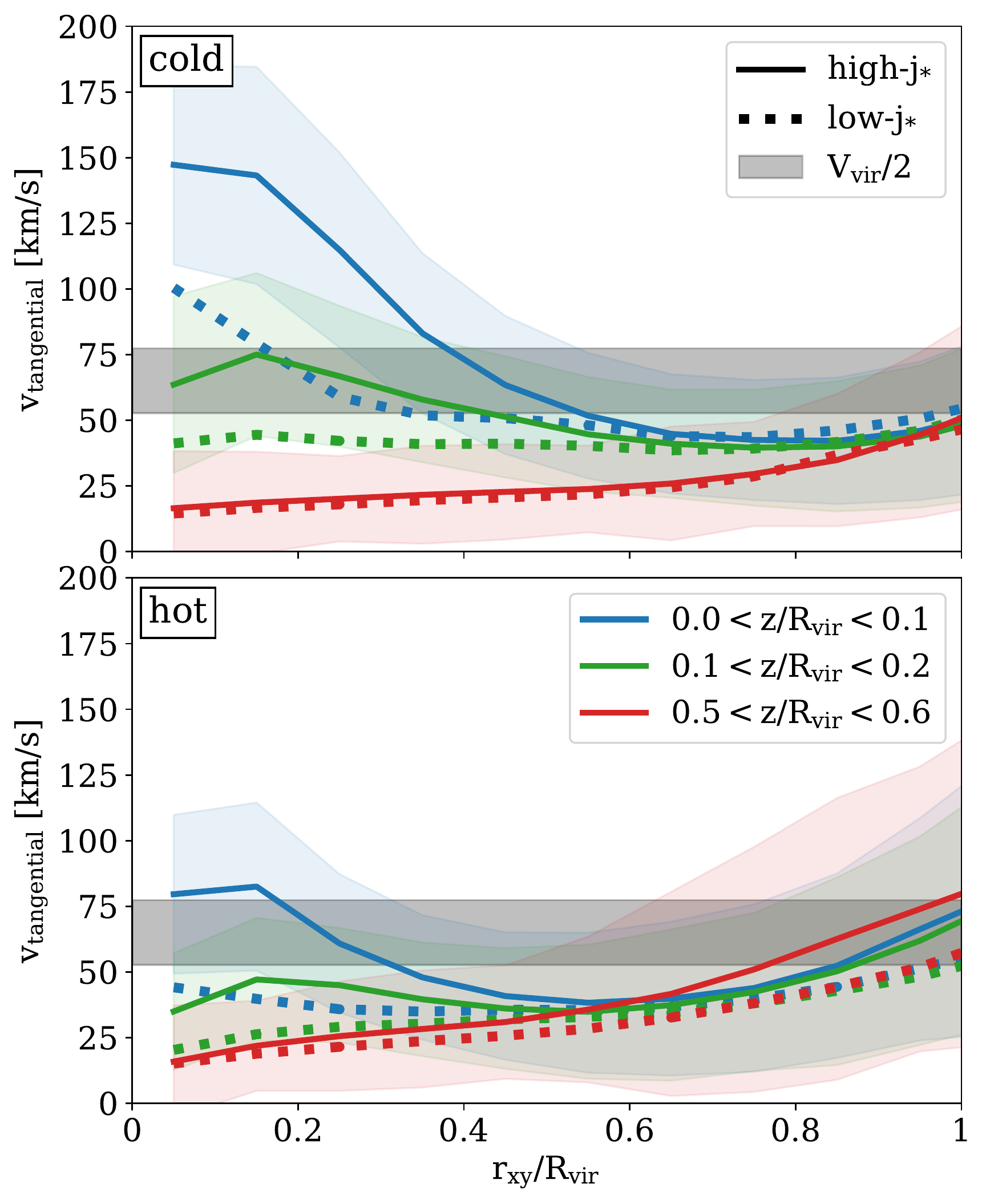}{0.48\textwidth}{}
\caption{
Tangential velocity profiles for cold (top panel) and hot (bottom panel) gas in the CGM of TNG100 MW-mass halos at $z=0$, aligned to the total angular momentum vector of the CGM. Each panel shows the CGM of high-$j_{*}$ (solid) and low-$j_{*}$ (dotted) galaxies at three different heights (blue, green, red). The colored shaded regions are the $\pm1\sigma$ scatter for the high-$j_{*}$ profiles and are of comparable size to those of of the low-$j_{*}$ profiles. The black shaded region shows the range of $V_{\rm vir}/2$ for all halos, where $V_{\rm vir} = \sqrt{GM_{\rm vir}/R_{\rm vir}}$ is the virial velocity.
}
\vspace{0.3cm}
\label{f:vprofile}
\end{figure}

We now seek to quantify the magnitude of the tangential (i.e. non-radial) velocities that contribute to the angular momentum magnitudes displayed so far. In Figure~\ref{f:vprofile}, we plot mass-weighted radial profiles (in cylindrical shells) of the vector-summed spherical tangential velocity of the cold and hot components of the CGM. As we are interested here in the motion within the CGM, we align these profiles to the CGM's total angular momentum vector (as in the rightmost columns of Figures \ref{f:coldjprofile} and \ref{f:hotjprofile}). Note that this alignment choice does not affect the computation of the tangential velocity but merely the location at which that velocity is displayed in the profiles. For radii $\lesssim 0.5 R_{\rm{vir}}$, we see that cold gas and gas around high-$j_{*}$ galaxies have higher tangential velocities than hot gas and gas around low-$j_{*}$ galaxies, respectively. At larger radii however, the high-$j_{*}$ and low-$j_{*}$ samples are nearly identical, and the hot gas has slightly higher velocities. For both cold and hot components, the largest tangential velocities occur close to the galactic plane and within $\sim 0.5 R_{\rm vir}$, where they can exceed half the virial velocity of the halo. The tangential velocities fairly quickly drop off with height and radius, reaching a value of $\approx 25 \; \rm{km} \; \rm{s}^{-1}$, independent of temperature, before they start to increase again in the outer halo, mostly for the hot gas. 

The tangential velocity as we have defined it contains rotation in two coordinates: the azimuthal coordinate $\phi$ and the polar coordinate $\theta$. We find (but do not show) that the shapes of the azimuthal velocity ($v_{\phi}$) profiles are almost identical to those of the tangential velocity profiles in Figure \ref{f:vprofile} but reduced by $\sim 10-20 \; \rm{km} \; \rm{s^{-1}}$. However, the average polar velocity ($v_{\theta}$) at all locations is $0 \; \rm{km} \; \rm{s^{-1}}$, meaning there is no coherent rotation in the $\theta$ direction at all. These results confirm that the high-angular momentum ``wedge'' we see in Figures \ref{f:coldjprofile} and \ref{f:hotjprofile} is associated with rotational velocities primarily in the azimuthal direction, which are a significant fraction of the virial velocity of the halo in the inner CGM.

\subsection{MW-mass Halos at $z > 0$} \label{sec:results2}

\begin{figure}
\fig{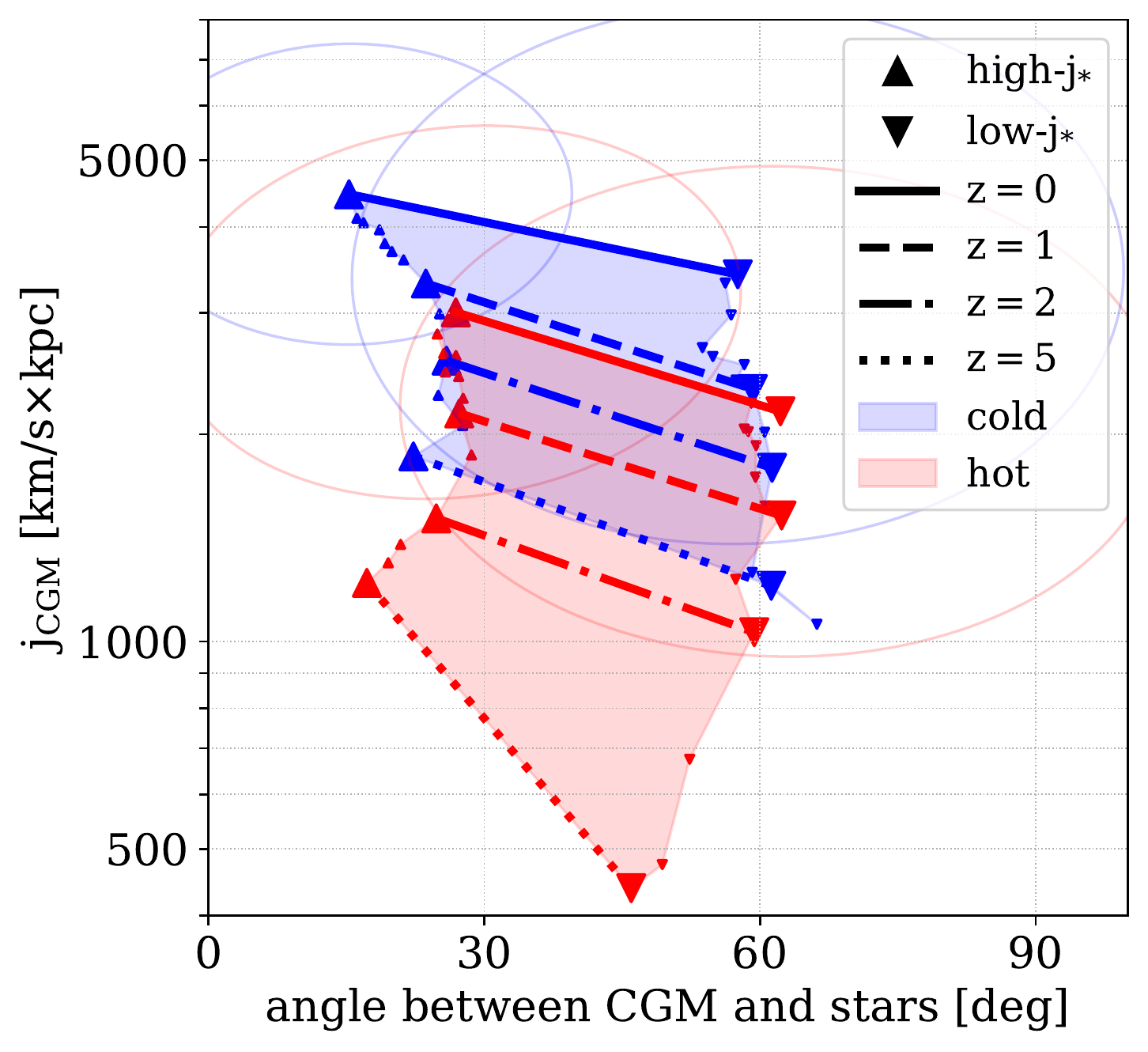}{0.48\textwidth}{}
\caption{
Specific angular momentum magnitude vs. misalignment to the stars for cold (blue) and hot (red) gas in high-$j_{*}$ (upward pointing triangles) and low-$j_{*}$ (downward pointing triangles) TNG100 MW-mass halos at $z=0$ to $z=5$. Ellipses show $1\sigma$ scatter of $z=0$ galaxies (identical to those in Figure \ref{f:totalj}) and are of similar size at all redshifts. Opaque lines connect high-$j_{*}$ and low-$j_{*}$ points at four selected redshifts and demonstrate the persistence of the misalignment angle and magnitude difference over cosmic time.
}
\vspace{0.3cm}
\label{f:totaljhigherz}
\end{figure}

In this section, we extend our analysis to halos that have a $z=0$ MW-mass at higher redshifts (and thus, a larger halo mass at $z=0$) by applying the same galaxy selection criteria as described in Section \ref{sec:methods} to redshifts $z > 0$. First, in Figure~\ref{f:totaljhigherz}, we plot the same quantities as shown in Figure~\ref{f:totalj} but at redshifts from $z=0$ to $z=5$ for the hot and cold smooth components of the CGM separately. The thick opaque lines connecting the median high-$j_{*}$ and low-$j_{*}$ points at different redshifts have similar slopes and lengths, both for hot gas and cold gas, thus indicating a remarkable consistency between magnitudes and alignments of high-$j_{*}$ and low-$j_{*}$ galaxies with redshift. The only significant difference with redshift is the total magnitude of a given component of the gas, which increases over time at approximately similar rates for the upper and lower quartiles of the stellar specific angular momentum distribution. This is consistent with the expected growth of specific angular momentum with redshift at fixed mass\footnote{Assuming a constant halo spin parameter $\lambda$: $j \propto \lambda R V \propto R^{1/2} \propto 1/(1+z)^{1/2}$.}: $j \propto (1+z)^{-1/2}$. There also appears to be a steady trend of decreasing misalignment angle for the cold CGM of high-$j_{*}$ galaxies at $z \lesssim 1$. However, broadly speaking, the total angular momentum properties of MW-mass halos are redshift independent, apart from the total angular momentum magnitudes themselves.

\begin{figure*}
\fig{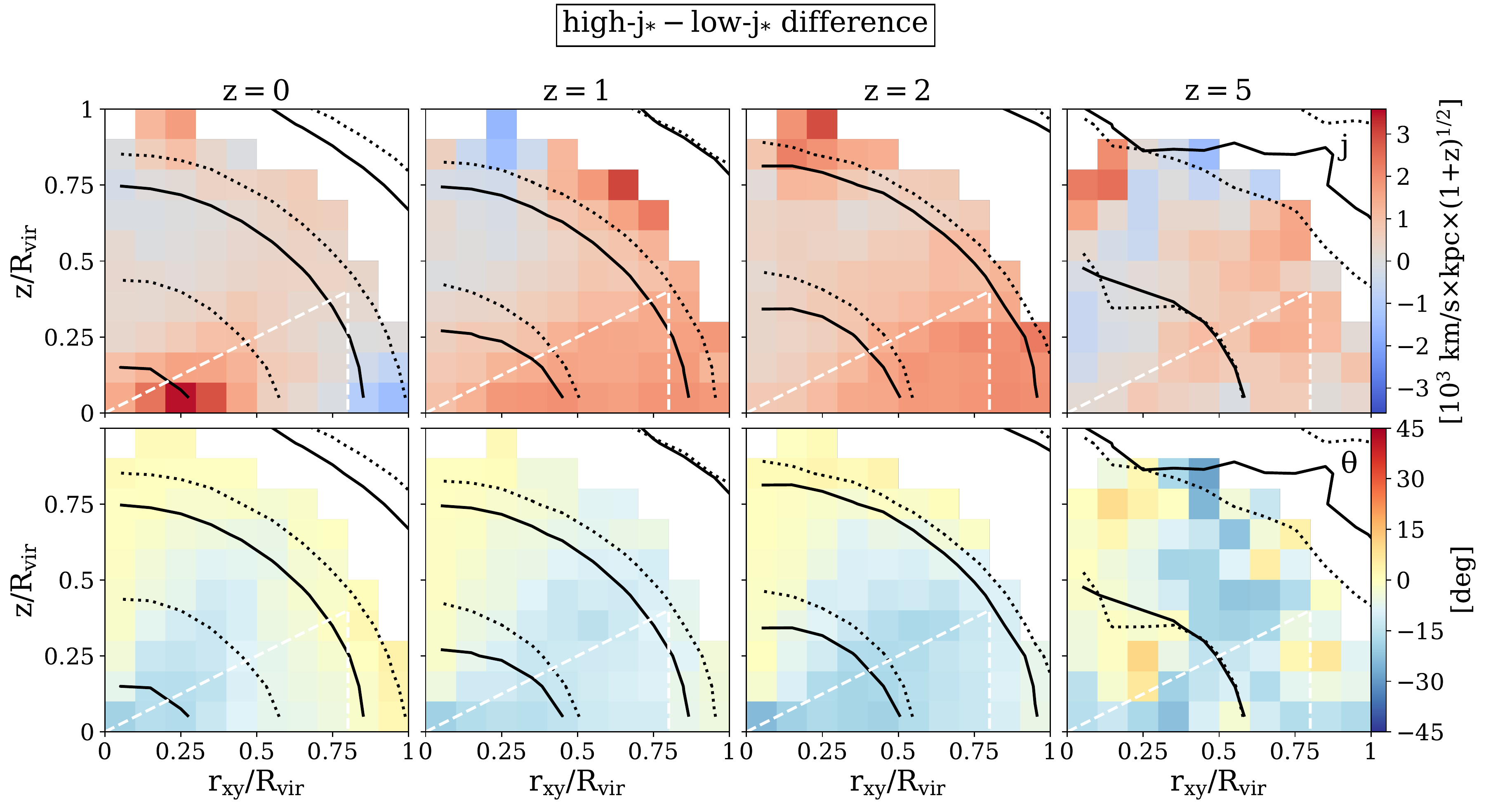}{0.8\textwidth}{}
\caption{
The difference between the mass-weighted spatial distributions of the cold CGM of high-$j_{*}$ and low-$j_{*}$ galaxies binned in height ($y$-axis) and cylindrical radius ($x$-axis; the same panels as the fourth column of Figure \ref{f:coldjprofile}) for TNG100 MW-mass halos at $z=0, \; 1, \; 2$, and $5$, from left to right. The scale of the first row is normalized by $(1+z)^{-1/2}$ to account for overall growth of angular momentum with redshift. The corresponding evolution of the hot CGM is very similar.
}
\label{f:coldjprofilehigherz}
\end{figure*}

In Figure \ref{f:coldjprofilehigherz}, we show the angular momentum magnitude and misalignment angle difference maps for cold gas, analogous to those in the fourth column of Figure \ref{f:coldjprofile} (i.e. aligned to the total angular momentum vector of the CGM) but for higher redshifts; to ease comparison, the first column of Figure \ref{f:coldjprofilehigherz} is identical to the fourth column of Figure \ref{f:coldjprofile}. We also normalize the angular momentum magnitude difference by $(1+z)^{-1/2}$ to remove the overall angular momentum growth from the difference plots. We immediately see that there are two key structural differences between the high-$j_{*}$ and low-$j_{*}$ galaxies that are redshift dependent. First, at very high redshifts ($z=5$), the magnitude and misalignment angle difference structure is much noisier than at all other redshifts, indicating that the organized structure seen at $z=0$ is the result of longer-term evolution. However, by $z=2$, the basic aligned ``wedge'' structure is in place. Second, at redshifts $z < 2$, the area of the strongest magnitude difference seems to drift inward toward the galaxy from $\sim R_{\rm vir}$ at $z=1-2$ to $\sim R_{\rm vir}/3$ at $z=0$. This is accompanied by the inward drift of the largest radius where the misalignment angle difference is negative, indicating a better intrinsic alignment around high-$j_{*}$ galaxies. In examining the actual magnitude and misalignment angle values as a function of redshift, we find that for the cold gas, the magnitude of $\mathbf{j}_{\rm{CGM}}$ increases in the outer parts of the halo over time, but only in the CGM of high-$j_{*}$ galaxies does the magnitude increase in the inner part of the halo, and this happens most dramatically after $z=1$. Neither the high-$j_{*}$ nor low-$j_{*}$ misalignment angle profiles change significantly after $z=2$. We also find the same features in the corresponding plots for hot gas, though the size of the magnitude effect is decreased. This potentially highlights a point in time at which angular momentum exchange between the galaxy and the CGM becomes particularly effective, presumably due to the emergence of fountain flows, and high-$j$ gas can exist nearer to the galaxy. 

\subsection{Other Halo Masses} \label{sec:results3}

\begin{figure}
\fig{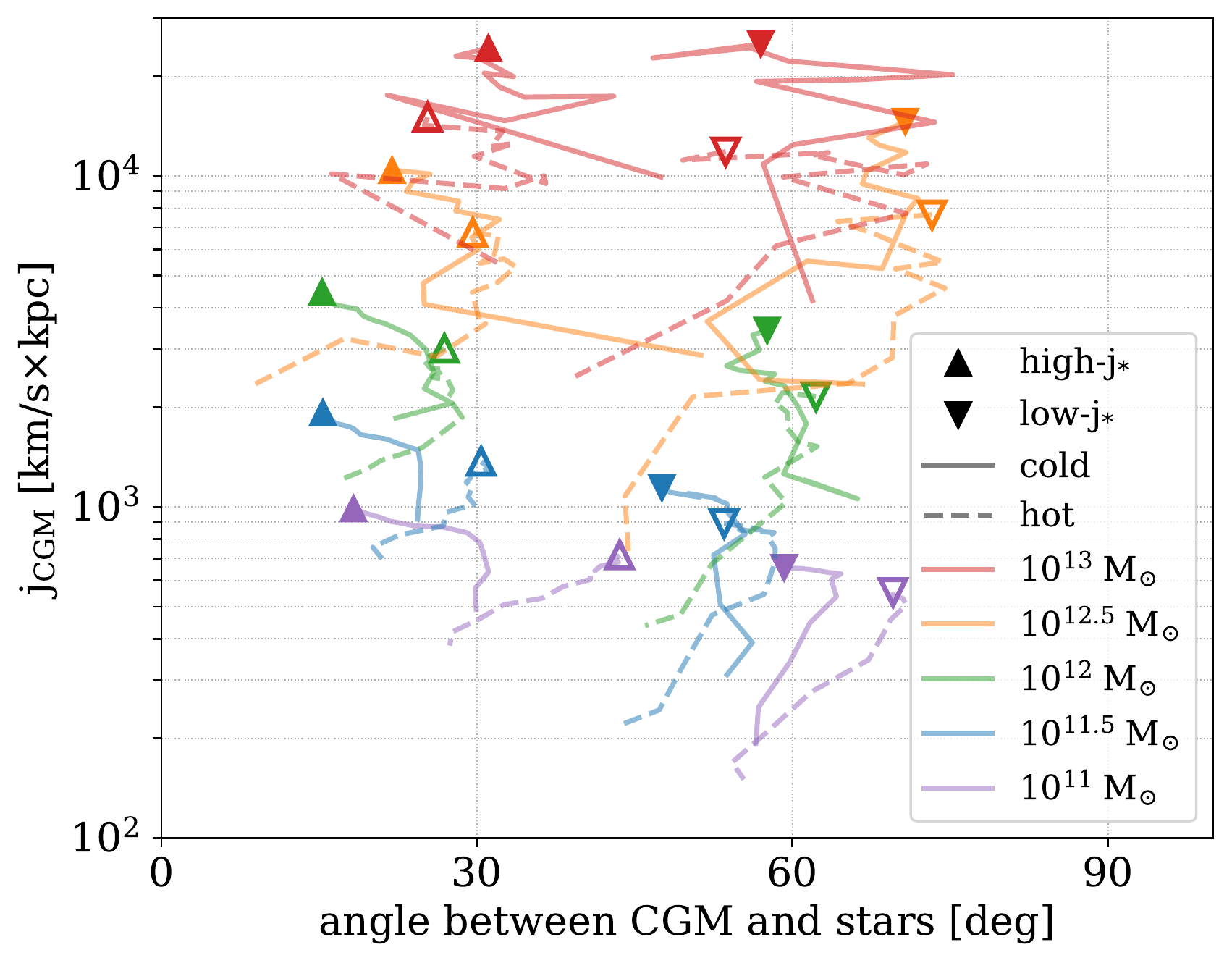}{0.48\textwidth}{}
\caption{
Specific angular momentum magnitude vs. stellar misalignment angle for cold (solid, filled triangles) and hot (dashed, empty triangles) gas in TNG100 halos of over two orders of magnitude in mass binned in five bins, each a different color. The triangles show $z=0$ values, and the lines show the population evolution within the bin up to $z=5$. 
}
\vspace{0.3cm}
\label{f:totaljall}
\end{figure}

Next, we consider the role of halo mass in influencing the angular momentum structure of the CGM. Figure~\ref{f:totaljall} displays the same cold and hot gas evolutionary tracks as shown in Figure~\ref{f:totaljhigherz} but now for five halo mass bins of width 0.5 dex centered on $10^{11} \; M_{\odot}$, $10^{11.5} \; M_{\odot}$, $10^{12} \; M_{\odot}$, $10^{12.5} \; M_{\odot}$, and $10^{13} \; M_{\odot}$. Clearly, higher-mass halos have more angular momentum, consistent with the expected scaling of $j \propto M_{\rm{vir}}^{2/3}$, but the misalignment angle seems to be essentially independent of halo mass. As in Figure \ref{f:totaljhigherz}, the misalignment of the cold CGM of high-$j_{*}$ galaxies decreases at $z\lesssim1$. We can also see evidence of two regimes of halo mass: one is $M_{\rm halo} \lesssim 10^{12} \; M_{\odot}$ (green, blue, and purple), for which all halos have a CGM with a higher-angular momentum magnitude and better alignment around high-$j_{*}$ galaxies compared to low-$j_{*}$ ones. The other regime is $M_{\rm{halo}} > 10^{12} \; M_{\odot}$ (orange, red). In these halos, the CGM angular momentum magnitude difference between high-$j_{*}$ and low-$j_{*}$ galaxies is consistent with zero, though the misalignment angle difference remains. The evolutionary tracks are also considerably more jagged. Additionally, for all galaxies, the misalignment difference between cold and hot gas increases toward lower halo masses at $z=0$, possibly demonstrating less mixing between the phases due to the lower ``hot'' (with respect to $T_{\rm{vir}}$) gas mass fraction in the CGM of those lowest-mass halos ($\approx 20\%$) compared to the highest-mass ones ($\approx 50\%$).

\begin{figure*}
\fig{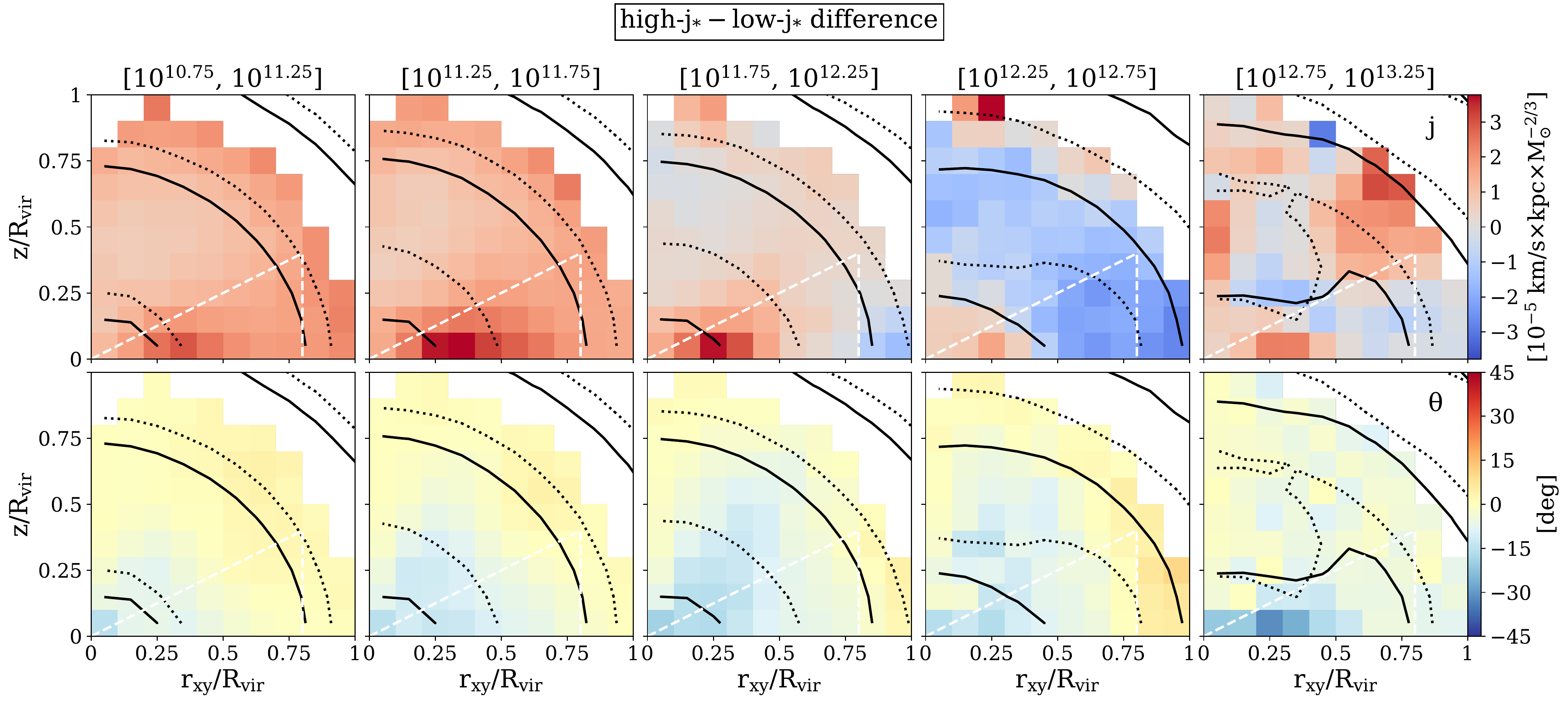}{\textwidth}{}
\caption{
The difference between the mass-weighted spatial distributions of the cold CGM of high-$j_{*}$ and low-$j_{*}$ galaxies binned in height ($y$-axis) and cylindrical radius ($x$-axis; the same panels as the fourth column of Figure \ref{f:coldjprofile}) for the cold CGM of five different TNG100 halo mass bins at $z=0$, which are displayed on the top of each column in units of $M_{\odot}$. The scale of the first row is normalized by $M_{\rm vir}^{2/3}$ to account for the median halo mass in each bin. The corresponding plots for the hot CGM differ slightly in the rightmost column but are otherwise very similar.
}
\label{f:coldjprofileotherm}
\end{figure*}

In Figure \ref{f:coldjprofileotherm}, we show cold gas difference maps at $z=0$ for the same five halo mass bins as in Figure~\ref{f:totaljall}, again, aligned to the total angular momentum vector of the CGM. First and foremost, we can clearly see the distinction between the first three columns and the last two: the lower-mass high-$j_{*}$ halos have an excess angular momentum in the wedge defined earlier compared to the low-$j_{*}$ halos and are similarly self-aligned. In the two highest-mass bins, the difference between the high-$j_{*}$ and low-$j_{*}$ is less clear. In examining each population by itself, we find that both contain a similarly sized excess in the wedge, but the structure outside the wedge is more complicated, sometimes resulting in an excess around the low-$j_{*}$ galaxies (fourth column) and sometimes little organized structure at all (fifth column). The misalignment angle maps do not vary much with mass, consistent with the overall result from Figure~\ref{f:totaljall}. We find broadly similar results when we examine the hot gas (not shown), though the difference map in the largest halo mass bin shows a nearly uniform slight angular momentum magnitude excess and a misalignment angle difference of nearly $0^{\circ}$. This is presumably due to two related factors: (1) the dominance of black hole feedback over stellar feedback at higher halo masses, and (2) the resulting dearth of typical high-$j_{*}$ spirals at halo masses $\sim 10^{13} \; M_{\odot}$ that renders the high-$j_{*}$/low-$j_{*}$ split of the galaxy population less physically meaningful \citep[see, for example, Figure 2 of][which shows this for the original Illustris simulation]{Genel15}. What does seem clear though is that the angular momentum structure of MW-mass halos (which we found to be largely independent of redshift) is not unique to MW-mass halos and represents a typical structure for a wide range of halos up to $\sim10^{12} \; M_{\odot}$.

\section{Discussion} \label{sec:discussion}

Having established the basic differences of the CGM angular momentum content between high- and low-$j_{*}$ galaxies, we now explore what coherent velocities in the CGM and variations to the IllustrisTNG physics model can tell us about the source of these differences. We then place the results in this paper in the larger context of previous CGM-related angular momentum results.

\subsection{Radial and Total Velocities}

\begin{figure}
\fig{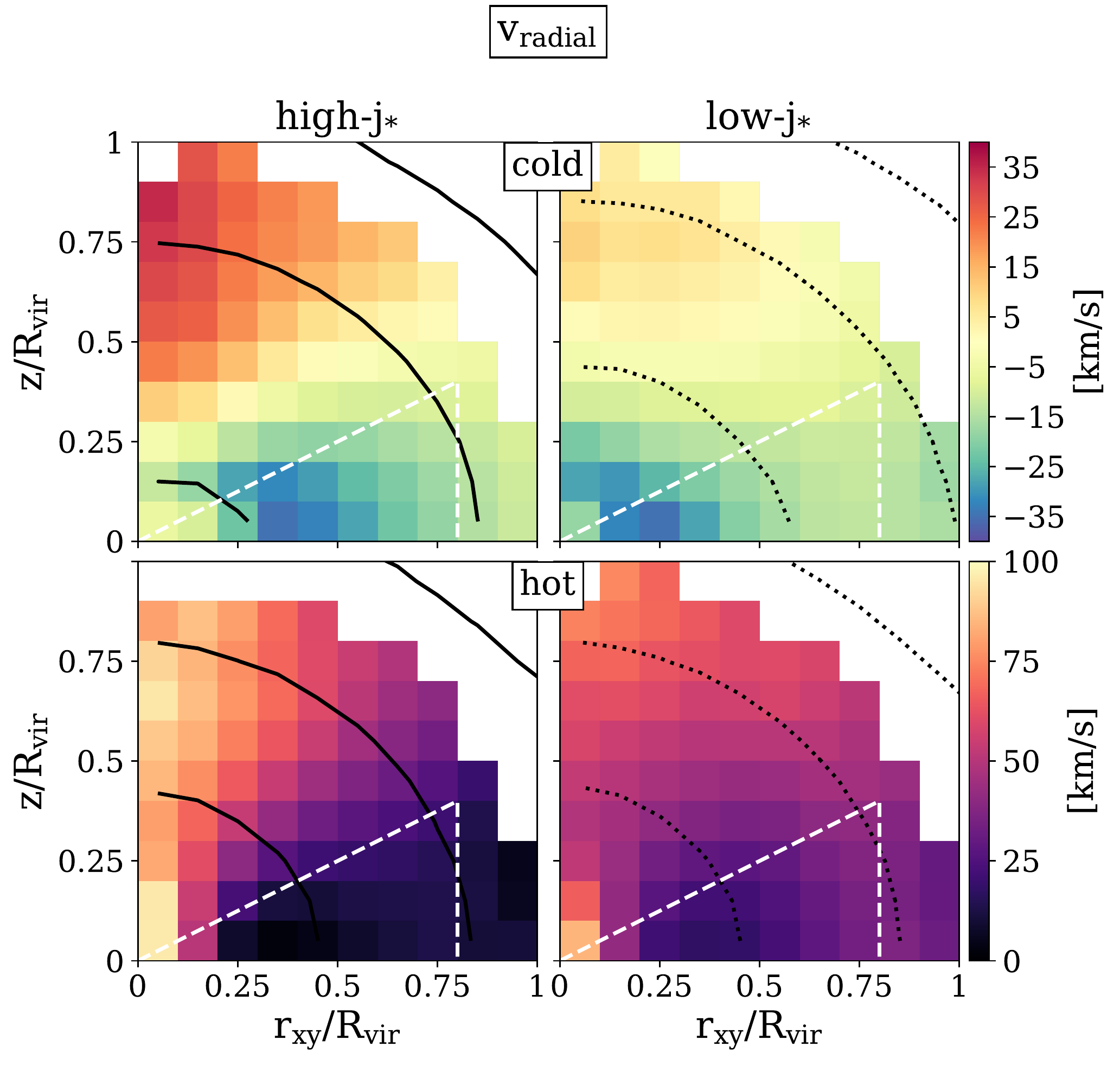}{0.48\textwidth}{}
\caption{
Mass-weighted spatial distributions of the radial velocity of the cold (top row panels) and hot (bottom row panels) CGM of TNG100 MW-mass halos at $z=0$, split into high-$j_{*}$ (left panels) and low-$j_{*}$ (right panels) populations as before. Mass contours are the same as the corresponding contours in Figures~\ref{f:coldjprofile} and \ref{f:hotjprofile}.
} 
\vspace{0.3cm}
\label{f:vradprofile}
\end{figure}

In this section, we identify a clear distinction between hot and cold gas, which so far have appeared dynamically very similar, i.e., the distribution of radial velocities. In Figure~\ref{f:vradprofile} we show radial velocity maps of the cold CGM (top two panels) and the hot CGM (bottom two panels) for our main sample of interest (MW-mass halos at $z=0$). For both high-$j_{*}$ (left panels) and low-$j_{*}$ (right panels) galaxies, the hot CGM is characterized by strongly outflowing gas in the polar regions and relatively weakly outflowing gas elsewhere. However, the cold gas shows a key difference between the two populations. While there is always inflowing cold gas at small heights above the galaxy as well as in the disk plane, only the CGM of high-$j_{*}$ galaxies shows the presence of net outflowing cold gas in the polar region of the CGM. 

This pattern of radial velocities is strongly suggestive of galactic fountains where gas in the CGM is continuously recycled, and it could potentially explain why the CGM of high-$j_{*}$ galaxies is so much more aligned to their galaxies than are the CGM of low-$j_{*}$ galaxies. We have found in previous work \citep{DeFelippis17} that the baryons locked in $z=0$ stars in high-angular momentum disks (comparable to the high-$j_{*}$ population in this paper) spend a significant amount of time participating in galactic fountains, while stars in galaxies simulated without feedback end up with a few times lower angular momentum. We also found in that work that participation in the fountains increases the specific angular momentum of the gas, meaning the outflowing/inflowing pattern in the cold gas may be broadly tracing the angular momentum growth in the CGM. 

Slightly more subtly, the regions of the strongest radial inflow also differ between the CGM of high-$j_{*}$ and low-$j_{*}$ galaxies: the former has essentially no radial motion in the plane of the galaxy at $<0.2 \; R_{\rm vir}$, possibly indicative of the much stronger rotation there (see Figure \ref{f:vprofile}). Evidently, low-$j_{*}$ galaxies are associated with stronger net radial inflows in their inner CGM. Further analysis of the mass (and angular momentum) participating in these radial flows requires the use of tracer particles, which we defer to a later work.

\begin{figure}
\fig{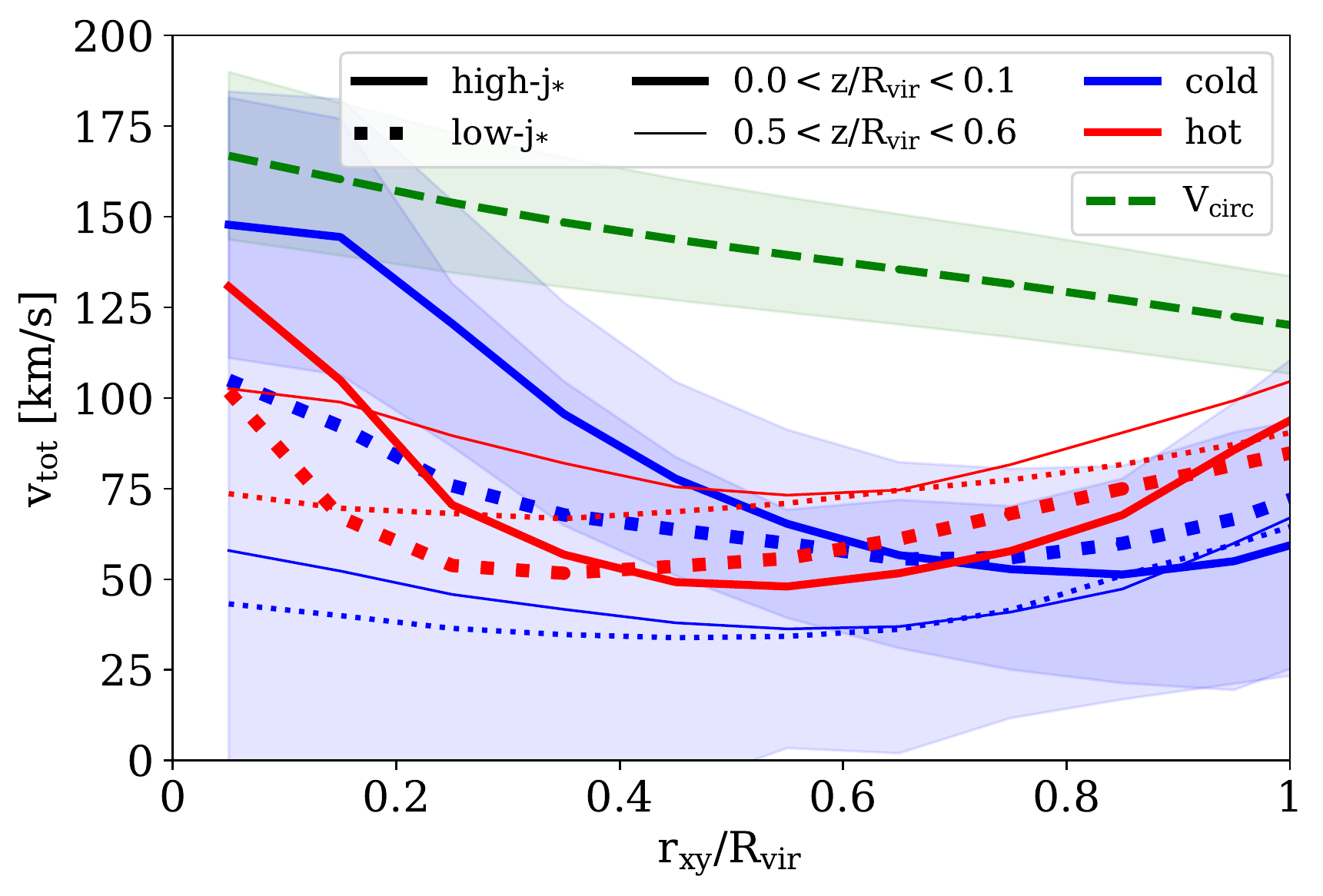}{0.48\textwidth}{}
\caption{
Total coherent velocity profiles for the cold (blue) and hot (red) CGM of TNG100 MW-mass halos at $z=0$, split into high-$j_{*}$ (solid) and low-$j_{*}$ (dotted) populations as before. We show profiles at a small (thick) and large (thin) height in the halo. The green dashed line is the circular velocity, $V_{\rm{circ}} = \sqrt{GM(<r)/r}$, where $M(<r)$ is the total mass enclosed in the spherical radius $r$. The colored shaded regions are the $\pm1\sigma$ scatter for the two cold high-$j_{*}$ profiles and are of comparable size for all profiles at a given height. 
} 
\vspace{0.3cm}
\label{f:vtotprofile}
\end{figure}

In Figure \ref{f:vtotprofile}, we estimate the extent to which the gas is kinematically supported by coherent motion by plotting the average total ``coherent'' velocity, defined as $\sqrt{v_{\rm{tan}}^{2}+v_{\rm{rad}}^{2}}$, where $v_{\rm{tan}}$ is the tangential velocity shown in Figure \ref{f:vprofile} and $v_{\rm{rad}}$ is the radial velocity shown in Figure \ref{f:vradprofile}. We see that total coherent velocity is always less than the circular velocity as a function of radius, except for cold gas within $\sim 0.1-0.2 \; R_{\rm{vir}}$ of high-$j_{*}$ galaxies, indicating that the vast majority of the CGM is not completely kinematically supported by coherent motion. The remainder of the support must come from a combination of random motion (i.e. velocity dispersion) and pressure; however, the precise measurement of these other factors is outside of the scope of this paper.

\subsection{Model Variations}

\begin{figure*}
\fig{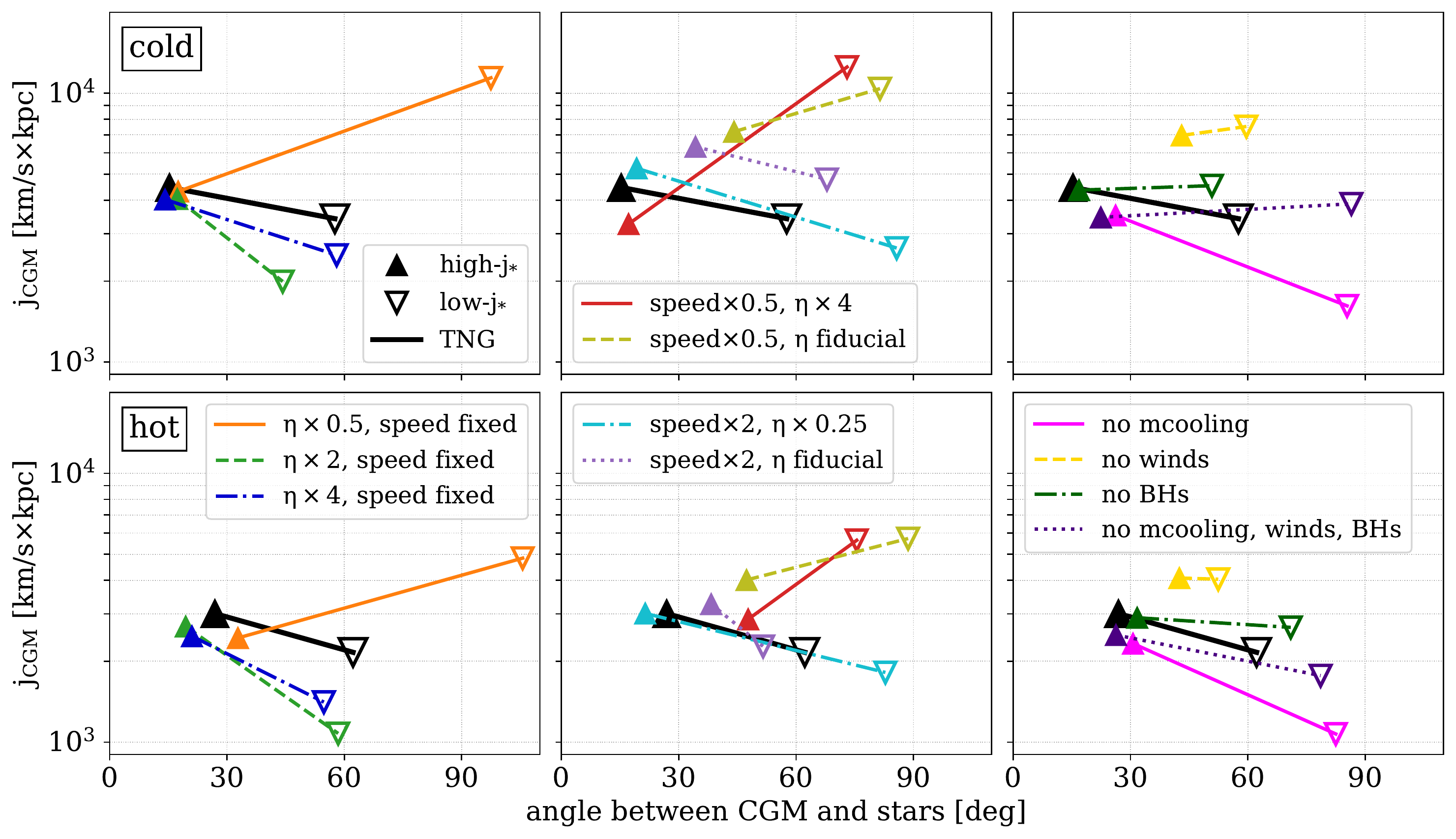}{0.8\textwidth}{}
\caption{
Specific angular momentum magnitude vs. misalignment to the stars of cold gas (top panels) and hot gas (bottom panels) in the CGM of MW-mass halos at $z=0$ for variations of the IllustrisTNG feedback model. The colored lines connect the high-$j_{*}$ (filled triangles) and low-$j_{*}$ (empty triangles) populations of each variation and emphasize that the qualitative misalignment angle difference between those two populations in the full TNG model (black) is also found in all other variations, with or without feedback.
}
\label{f:variations}
\end{figure*}

Next, we use some of the IllustrisTNG model variations described in \cite{Pillepich18a} (each one a $\approx (37 \; \rm{Mpc})^{3}$ box with a baryonic mass resolution of $2.4\times10^{6} \; M_{\odot}$ per cell, comparable to TNG100) to investigate the sensitivity of the angular momentum of the CGM to changes in the IllustrisTNG physics model. These changes are summarized in Figure~\ref{f:variations}, which shows the same quantities as in Figures~\ref{f:totalj}, \ref{f:totaljhigherz}, and \ref{f:totaljall} (i.e. CGM angular momentum magnitude and misalignment angle with respect to the stars). We focus on two types of variations: (1) those that change a property of the galactic wind, the dominant form of feedback for MW-mass halos, and (2) those that remove one or more physical processes completely. In each simulation, we perform the same halo selection, specific angular momentum cut, and temperature separation as described in Section~\ref{sec:methods}. 

Variations of the wind model change at least one of two quantities: the speed of the wind, and the mass loading ($\eta$) of the wind. The left column of Figure~\ref{f:variations} shows the total magnitude and misalignment angle of three simulations with varying values of $\eta$ and fixed wind speed, both for cold (top panels) and hot gas (bottom panels). We see that increasing the mass loading does not change the angular momentum of the CGM very much compared to the fiducial properties, but reducing the mass loading drastically changes the properties around low-$j_{*}$ galaxies by increasing their CGM's angular momentum to be greater than those around high-$j_{*}$ galaxies and significantly worsening their alignment. The middle column panels show simulations where the wind speed is varied, and the mass loading is either kept at the fiducial value or varied in tandem with the wind speed so as to keep a fixed specific kinetic wind energy. The simulations with increased wind speed (whether or not $\eta$ is kept fixed) are also qualitatively similar to the fiducial model, while those with decreased wind speed have larger misalignment angles and higher-angular momentum magnitudes around low-$j_{*}$ galaxies.

Taken together, we find that the angular momentum magnitude of the CGM, and to a lesser extent, the misalignment angle, is sensitive to the strength of the wind but mostly only if the wind is ``weaker'' than the fiducial model. Stronger winds do not significantly change the angular momentum of the CGM. These conclusions apply to both the cold and the hot phases of the CGM. It is important to note, however, that the difference in misalignment angle between high-$j_{*}$ and low-$j_{*}$ galaxies is a consistent feature of all of the simulations. A detailed or quantitative interpretation of these sensitivities is difficult to achieve (see also \citealp{Pillepich18a}), but we hypothesize in a general sense that the reason that weaker winds lead to worse galaxy-CGM alignments and to higher CGM specific angular momentum magnitudes is that the diminished wind feedback allows for more low-angular momentum gas to form stars and stay locked in the galaxy rather than return to the CGM and thereby lower the CGM angular momentum and ``mix'' it with that of the galaxy's.

We also checked angular momentum properties for simulations with various aspects of the IllustrisTNG model removed completely: specifically, with no galactic winds, no metal cooling, and no black holes. We also checked a simulation that had all three of these mechanisms removed. The results, shown in the right column panels of Figure \ref{f:variations}, indicate that removing only the winds has the greatest effect on the CGM's angular momentum by increasing its magnitude around all galaxies; removing black holes has negligible effects on the magnitude, but removing metal cooling slightly lowers the magnitude. In all cases, the misalignment angle difference between high-$j_{*}$ and low-$j_{*}$ galaxies remains. Interestingly, removing all three forms of feedback at once resulted in changes to the fiducial model smaller than when only one form was removed. 

There are many caveats to this analysis that make interpretation difficult. As demonstrated previously \citep{Pillepich18a}, different forms of feedback often interact with each other in nonlinear ways, which is evident here in the third column of Figure~\ref{f:variations}: knowing how removing each individual form of feedback affects the CGM does not obviously inform how removing all three together affects the CGM. Additionally, any change to the fiducial model will in some way change galactic properties and could therefore affect the properties of galaxies classified as high-$j_{*}$ or low-$j_{*}$, not just their CGM. Other studies comparing simulations with different subgrid models \citep{Kauffmann19} and including different physics such as cosmic-rays \citep{Buck19} further demonstrate how sensitive properties of the CGM can be to the strength and implementation of various forms of feedback. Nevertheless, there are three clear conclusions we can draw from this analysis. First, median $j$ and $\theta$ values for the CGM depend somewhat on the properties of the wind and the presence of feedback. Second, the high-$j_{*}-$low-$j_{*}$ difference in angular momentum magnitude is usually positive but can flip sign if the feedback is weak enough. Finally, the high-$j_{*}-$low-$j_{*}$ difference in misalignment angle is always positive and, thus, is not driven by feedback.

\subsection{Comparisons to Previous Work}

We highlight here important results from other recent studies of the CGM and whether or not our results are consistent with them. The main finding by \cite{Stewart17} was that the presence of high specific angular momentum gas in the halo ($\sim4$ times larger than that of the dark matter) is independent of simulation code, suggesting that it is a fundamental characteristic of galaxy formation. Our results confirm the presence of such gas in a large population of galaxies (compared to a single halo studied by \citealp{Stewart17}), and we furthermore show that high-angular momentum galaxies have the highest angular momentum gas in their CGM. They also found that large-scale filamentary inflows resembling an extended cold disk can form around MW-mass galaxies (strongly resembling the spatial pattern of cold gas we find in Figure~\ref{f:coldjprofile}) and that velocities of such gas can be as large as $1.5\times$ the virial velocity of the halo ($\sim250 \; \rm{km} \; \rm{s}^{-1}$). The tangential velocities of cold gas in TNG100 MW-mass halos are not generally that high (see top panel of Figure \ref{f:vprofile}) but can exceed the virial velocity close to the disk. Other studies of simulated galaxies at high redshifts \cite[e.g.][]{Kimm11,Danovich15} have found that inflowing cold gas streams can transport angular momentum through the halo toward the galaxy while maintaining its high spin with respect to the dark matter. While we only look at average velocity structures in TNG100 and do not follow individual gas streams, the radial velocity maps (Figure~\ref{f:vradprofile}) nevertheless show that much of the cold rotating (and high-$j$) gas in the CGM has a net inflowing velocity out to nearly the virial radius at $z=0$, perhaps suggesting it is indeed a source of the baryonic angular momentum of the galaxy.  

Recent work with EAGLE, another modern large-scale cosmological simulation, has also found evidence of rotation in the CGM. \cite{Oppenheimer18} measured the spin parameter of the hot halos of $L*$ galaxies and found that they are comparable to those of the cold gas, even though the hot gas is more spherically distributed. We also find a similar relationship between the cold and hot CGM and provide further support to the \cite{Oppenheimer18} conclusion that rotation in the CGM, especially in the inner parts, is a significant deviation from hydrostatic equilibrium that models of the CGM should take into account. \cite{Ho19} focus on the cold gas around a single EAGLE galaxy in a MW-mass halo and find corotating gas that would be detectable observationally at low azimuthal angles from the galaxy ($\lesssim 10^{\circ}$) and impact parameters $\lesssim 60 \; \rm{kpc}$ ($\sim 0.2-0.3 \; R_{\rm vir}$ for this halo mass). This is comparable to the region of the CGM in TNG100 MW-mass halos where cold gas is rotating near or above the virial velocity (see Figure~\ref{f:vprofile}). \cite{Ho19} also measure typical inflow speeds between $20$ and $60 \; \rm{km} \; \rm{s}^{-1}$, which match fairly well with the net inflow we measure in TNG100. They further find that gas tends to accrete anisotropically in structures that extend further in cylindrical radius than height, which is, again, qualitatively matched by our radial velocity maps. As a whole, we find signs for good agreement between the rotational properties of the CGM between the EAGLE and TNG100 simulations. We further demonstrate in this work that the hot and cold CGM components have similar angular momentum properties as well, though the cold CGM always has a higher magnitude than the hot CGM. This supports results from \cite{Danovich15} who found that while cold and hot gas in their (smaller) sample of MW-mass halos at $z > 1$ have similarly shaped spin profiles, the cold gas has a factor of $\sim 2$ higher spin than the hot gas.

Rotation in the CGM is difficult to measure observationally, but recent efforts provide powerful evidence of its prevalence. \cite{MartinC19} use 50 pairs of galaxies and quasar sightlines to measure Mg II absorption and find that it is preferentially located along both the major axis of the galaxy, consistent with corotation, and the minor axis, indicative of biconical outflows. \cite{Zabl19} use a smaller sample taken with the MUSE instrument on the Very Large Telescope and find similar results in Mg II, as well as inferred accretion rates consistent with simulations and previous observations of Mg II \cite[e.g.][]{Bouche12,Kacprzak12,Bouche13}. \cite{Hodges-Kluck16} use X-ray measurements of O VII to detect a rotating hot halo around the MW (though with a large misalignment angle) that contains as much angular momentum as the stars. Taken all together, these observational results are at least qualitatively consistent with our measurements of the CGM's angular momentum in TNG100. There are discrepancies though, such as the rotational velocity inferred by \cite{Hodges-Kluck16} of $\sim180 \; \rm{km} \; \rm{s}^{-1}$, which is much larger than our measured hot gas rotational velocities in TNG100. However, a more careful comparison to observations is necessary to properly address this, which we defer to a future paper.

\section{Summary} \label{sec:summary}

We have calculated and characterized the angular momentum of the CGM in the TNG100 simulation. We focus on the smooth CGM, namely halo gas excluding the ISM (i.e. outside twice the stellar half-mass radius of the galaxy) as well as satellites, and in particular on $z=0$ MW-mass halos. Our main conclusions are as follows:

\begin{enumerate}
    \item The total specific angular momentum of the smooth CGM around galaxies with high stellar spin (high-$j_{*}$) is systematically larger and better aligned to the stellar body of the galaxy than that of the CGM around low-$j_{*}$ galaxies, both for hot and cold gas. The satellite component has a higher specific angular momentum but in general much less mass than the smooth component.
    
    \item High-angular momentum cold gas around high-$j_{*}$ galaxies is distributed in a large structure that is well aligned with the galaxy plane (defined as the plane perpendicular to the galaxy angular momentum vector) and has an opening angle of $\sim30^{\circ}$. Low-$j_{*}$ galaxies do not have such a structure in their CGM with respect to the galaxy. However, the spatial distributions of self-alignment of the cold CGM around high-$j_{*}$ and low-$j_{*}$ galaxies are very similar, indicating that the misalignment difference between the populations is largely due to an overall galaxy-CGM misalignment in the low-$j_{*}$ case, rather than internal structural differences between the CGM of the two types of galaxies. 
    
    \item The spatial angular momentum distribution of the hot CGM is not structurally different between the two galaxy populations, but the hot gas in the CGM of high-$j_{*}$ galaxies is systematically better aligned and has a higher magnitude throughout the halo. Furthermore, the inner half of the hot CGM around high-$j_{*}$ galaxies is dominated by rotation around the galaxy, but the outer half is dynamically very similar to the same area of the CGM of low-$j_{*}$ galaxies.
    
    \item These CGM characteristics are roughly independent of halo mass and redshift for halos with masses $\lesssim 10^{12} \; M_{\odot}$, but for halo masses $> 10^{12} \; M_{\odot}$, the high-$j_{*}-$low-$j_{*}$ difference in magnitude is no longer positive, and the difference maps are much noisier. This is likely due to the increased influence of AGN feedback affecting the properties of galaxies and halos at these masses.
    
    \item The CGM of high-$j_{*}$ galaxies contains outflowing and accreting cold (relative to $T_{\rm{vir}}$) gas characteristic of galactic fountains, whereas the CGM of low-$j_{*}$ galaxies has no significant cold outflows. This points to stronger gas mixing and, thus, a stronger dynamical connection between the galaxy and the CGM of high-$j_{*}$ galaxies.
    
    \item The precise form and parameters of the galactic wind feedback model can affect the angular momentum properties of the CGM, but the misalignment angle difference between the CGM of high-$j_{*}$ and low-$j_{*}$ galaxies is always present, suggesting its existence results from cosmic gas accretion and, thus, is a fundamental part of galaxy formation.

\end{enumerate}

We find our results to be qualitatively consistent with recent studies of rotation in the CGM, both for other simulations run with different subgrid models (EAGLE) and observations using quasar sightlines. Future work is necessary to elucidate the origins of the angular momentum trends we see and whether they are primarily set by cosmic inflows, feedback, or both in tandem. Nevertheless, the robustness of the trends across large ranges of halo mass and cosmic time emphasizes the important role the CGM's angular momentum has on galaxy evolution.

\acknowledgments
We thank Nicolas Bouch\'e, Mary Putman, Ari Maller, Jolanta Zjupa, and Drummond Fielding for insightful and useful discussions. We also thank the anonymous referee for their helpful comments. Support for program numbers HST-AR-14565 and HST-AR-15022 was provided through a grant from the Space Telescope Science Institute, which is operated by the Association of Universities for Research in Astronomy, Incorporated, under NASA contract NAS5-26555. D.D. acknowledges support from the Chateaubriand Fellowship Program. G.L.B. acknowledges support from NSF grants AST-1615955 and OAC-1835509 and NASA grant NNX15AB20G. The Flatiron Institute is supported by the Simons Foundation.

\vspace{5mm}

\software{\textsc{NumPy} \citep{vanderWalt11}, \textsc{Matplotlib} \citep{Hunter07}, and \textsc{IPython} \citep{Perez07}}

\bibliographystyle{aasjournal}
\bibliography{references}

\end{document}